%% file: paper.tex
\documentclass[twocolumn,letterpaper,10pt]{article}
\usepackage{tikz}
\usepackage{pgfplots}
\pgfplotsset{compat=newest}

\usepackage[margin=1in]{geometry}

\usepackage{tabularx,booktabs,multirow,threeparttable}
\usepackage{listings}
\usepackage{import}

\usepackage{hyperref}

\usepackage{amsmath,amsfonts}
\usepackage{subcaption}
\usepackage{comment}

\usepackage{siunitx}
\usepackage{tikz}
\usepackage{amsmath}
\usetikzlibrary{arrows,calc,automata,patterns}

\newcommand{\projectname}{VGF}
\newcommand{\vgfa}{VGF}
\newcommand{\vgfb}{VGF\_fast}

\newcommand{\orlando}[1]{{\color{blue} [orlando]: #1}}
 
\newcommand{\Ruochen}[1]{{\color{orange} [Ruochen]: #1}}

\date{}
\title{\Large\bf \projectname{}: Value-Guided Fuzzing\\-- Fuzzing Hardware as Hardware --}
\author{%
{\rm Ruochen Dai}\\
University of Florida
\and
{\rm Michael Lee}\\
University of Florida
\and
{\rm Patrick Hoey}\\
University of Massachusetts Lowell
\and
{\rm Weimin Fu}\\
Kansas State University
\and
{\rm Tuba Yavuz}\\
University of Florida
\and
{\rm Xiaolong Guo}\\
Kansas State University
\and
{\rm Shuo Wang}\\
University of Florida
\and
{\rm Dean Sullivan}\\
University of New Hampshire
\and
{\rm Orlando Arias}\\
University of Massachusetts Lowell
}

\import{styles/}{styles.tex}

\begin{document}

	\maketitle

	\begin{abstract}
		\import{abstract/}{abstract.tex}
	\end{abstract}

	\section{Introduction}
	\import{introduction/}{introduction.tex}\label{sec:introduction}



	\section{Reconciling a New Approach}
	\import{vgf/}{vgf.tex}

	\section{Prototyping}
	\import{prototype/}{prototype.tex}

	\section{Experimental Results}
	\import{results/}{results.tex}


    \section{Related Work}

\import{background/}{background.tex}

    \section{Conclusions}
    \import{conclusions/}{conclusions.tex}
        
	\bibliography{bib/references}
	\bibliographystyle{plain}

    \begin{appendix}
    \section{Appendix}

\import{discussion/}{discussion.tex}
    

    \end{appendix}
 
\end{document}

%% file: styles/styles.tex
\definecolor{CommentGreen}{rgb}{0,.6,0}
\definecolor{numbercolour}{gray}{0.5}
\definecolor{keywordc}{rgb}{.63,0,.42}
\definecolor{comment}{rgb}{0,.6,0}
\definecolor{keyword}{rgb}{.63,0,.42}
\definecolor{kw2}{rgb}{.50,.50,.15}
\definecolor{kw3}{rgb}{.42,.42,.63}
\definecolor{string}{rgb}{1,0,0}

\lstdefinestyle{vcode}{
	escapechar=@@,
	basicstyle=\ttfamily\footnotesize,
	language=Verilog,
	frame=tb,
	numbers=left,
	xleftmargin=\parindent,
	numbersep=2.5pt,
	fontadjust=true,
	basewidth=0.5em,
	showstringspaces=false,
	keywordstyle=\bfseries\color{magenta},
	commentstyle=\color{CommentGreen},
	tabsize=4,
	morekeywords={always_ff,always_comb,always_latch,priority,unique,%
	interface,modport,enum,struct,foreach,logic,typedef,bit,packed,unpacked,%
	tagged,union,endinterface,\$dumpfile,\$dumpvars%
	}
}
\lstdefinestyle{scode}{
	basicstyle=\ttfamily\footnotesize,
	language=Lisp,
	frame=tb,
	numbers=left,
	fontadjust=true,
	basewidth=0.5em,
	showstringspaces=false,
	keywordstyle=\bfseries\color{magenta},
	commentstyle=\color{CommentGreen},
	tabsize=4,
	morekeywords={if,define_insn,match_operand,include,unspec_volatile,restrict},
	xleftmargin=\parindent,
	numbersep=2.5pt
}
\lstdefinestyle{ccode}{
	basicstyle=\ttfamily\footnotesize,
	language=C++,
	frame=tb,
	numbers=left,
	xleftmargin=\parindent,
	numbersep=2.5pt,
	fontadjust=true,
	basewidth=0.5em,
	showstringspaces=false,
	keywordstyle=\bfseries\color{magenta},
	commentstyle=\color{CommentGreen},
	tabsize=4,
	morekeywords={mmap,mprotect,clone,futex,__asm__,uint32_t,override,constexpr,fork}
}

%% file: abstract/abstract.tex
As the complexity of logic designs increase, new avenues for testing digital hardware becomes necessary. Fuzz Testing (fuzzing) has recently received attention as a potential candidate for input vector generation on hardware designs. Using this technique, a fuzzer is used to generate an input to a logic design. Using a simulation engine, the logic design is given the generated stimulus and some metric of feedback is given to the fuzzer to aid in the input mutation. However, much like software fuzzing, hardware fuzzing uses code coverage as a metric to find new possible fuzzing paths. Unfortunately, as we show in this work, this coverage metric falls short of generic on some hardware designs where designers have taken a more direct approach at expressing a particular microarchitecture, or implementation, of the desired hardware.

With this work, we introduce a new coverage metric which employs not code coverage, but state coverage internal to a design. By observing changes in signals within the logic circuit under testing, we are able to explore the state space of the design and provide feedback to a fuzzer engine for input generation. Our approach, Value-Guided Fuzzing (VGF), provides a generic metric of coverage which can be applied to \emph{any} design regardless of its implementation. In this paper, we introduce our state-based VGF metric as well as a sample implementation which can be used with \emph{any} VPI, DPI, VHPI, or FLI compliant simulator, making it completely HDL agnostic. We demonstrate the generality of VGF and show how our sample implementation is capable of finding bugs considerably faster than previous approaches.


%% file: introduction/introduction.tex
Computing is trending toward the use of high-level languages supported by 
domain-specific platforms that incorporate specialized components to extract both 
performance and energy efficiency. This is in part due to the end of Dennard 
Scaling~\cite{hennessy2019new}, but also from a recognition that 
general-purpose machines are ill-suited to address current computing 
problems~\cite{liu2016cambricon, mao2022genpip} or are inherently 
insecure~\cite{lipp2020meltdown, kocher2020spectre}. Both industry and academia 
have invested significant effort into accelerating high-level 
payloads~\cite{jouppi2017datacenter, lie2022cerebras} or rapidly-prototyped 
security 
solutions~\cite{park2022diva, zhou2022ppmlac, samardzic2022craterlake}. 
FPGAs are currently being offered in nearly all major
datacenters~\cite{xilinx2022acceleration} and all major EDA 
companies offer cloud services~\cite{venkatachar2022eda}.
The latter is of particular interest because of the complex 
hardware design process using hardware description languages (HDL) such as 
VHDL or Verilog~\cite{ashenden2010designer}. 

Designing hardware using EDA tools typically involves 
iterative  improvements to meet space, time, or power constraints and heavily 
relies on built-in optimizations~\cite{gulati2010hardware, gayathri2017rtl} and
expertise. As such, increased exposure from non-experts to the hardware design process will likely be error prone at the least. This cause for alarm is justified by the clear correlation between number of lines of code, 
native/third-party library reliance, and number of users on the 
one hand and number of errors and/or reported vulnerabilities on the other~\cite{woody2014predicting, grimes2015beware}. This is evidenced by Common Weakness Enumeration
(CWE) reports~\cite{cweteam2023common} where 13 instances of weaknesses are indexed by \textit{verilog} 
and \textit{VHDL}, and 100 out of 933 examples of 
\textit{hardware design} weaknesses are delineated.
It's unsurprising, therefore, that recently significant effort in the hardware 
security community has been paid to the automated analysis of hardware 
(trojan) 
vulnerability detection~\cite{zhang2018end, hur2021difuzzrtl, 10.1145/3240765.3240842,
kande2022thehuzz, trippel2022fuzzing, muduli2020hyperfuzzing} by borrowing 
concepts from software fuzzing.

Hardware fuzzing approaches, to the best of our knowledge, currently employ control-flow guided 
feedback in one form or another to track code coverage and have successfully found many 
hardware bugs in designs with explicit control flow that allows instrumentation. However, control-flow information may not be obtainable based on the implementation of the design. Control-flow information, for example, can not be recovered from a gate-level netlist or a micro-coded design. This is true in general for ROM-table designs of which many exist. In fact, control-flow in a hardware design is only generated in certain instances, such as when explicit multiplexing or decoding logic are inferred using \lstinline[style=vcode]{if}/\lstinline[style=vcode]{else}, or \lstinline[style=vcode]{case} statements in the HDL description.

\noindent\textbf{An Illustrative Example.}
Consider a finite state machine (FSM) in which we need to navigate every state to trigger a bug, for example, a design in which the trigger signal is on the last state. When targeting ASICs, a hardware designer will explicitly write the hardware in a hardware description language (HDL) to fully describe its intent. We illustrate such an FSM in Listing \ref{lst:fsm_asic} using a horizontal microcode implementation.
The code makes use of an asynchronous ROM-table which holds the state transition as well as output of the FSM. 
\\

\begin{lstlisting}[style=vcode,caption={A finite state machine written in application-specific SystemVerilog HDL.},label={lst:fsm_asic}]
module fsm (input clk, res, d_in, output d_out);
    typedef struct packed {
            logic [1:0] next_state;
            logic       out;
        } ucode_line;
    typedef ucode_line ucode_table [0:7];
    logic [1:0] state;
    ucode_table ucode_rom = '{
            '{ next_state: 2'b00, out: 1'b0 },  /* 000 */
            '{ next_state: 2'b01, out: 1'b1 },  /* 001 */
            '{ next_state: 2'b01, out: 1'b0 },  /* 010 */
            '{ next_state: 2'b10, out: 1'b1 },  /* 011 */
            /* ... */
        };
    always @(posedge clk)
        if(res) state <= 2'b00;
        else state <= ucode_rom[{state, d_in}].next_state;
    assign d_out = ucode_rom[{state, d_in}].out;
endmodule
\end{lstlisting}



This microcoded FSM example above translated into C++ using Verilator~\cite{snyder2013verilator} uses bit-shifts and lookups into an array, Listing~\ref{lst:verilator_ucode}. No explicit control flow information can be obtained and, hence, no control-flow guided coverage can be inferred. Simply put, a coverage-guided fuzzer can not instrument a micro-code ROM's control-flow. It, therefore, can not dynamically determine what inputs drive a transition from one state to another state in the FSM, which means it can not learn a preferential input to get to the buggy state. Essentially, this reduces the coverage-guided fuzzer to random search thereby undermining its purpose. 

\begin{lstlisting}[style=ccode,caption={Fragment of Verilator generated code for microcoded FSM design.},label={lst:verilator_ucode}]
    // ...
    vlSelf->fsm__DOT__ucode_rom[0U] = 0U;
    vlSelf->fsm__DOT__ucode_rom[1U] = 3U;
    vlSelf->fsm__DOT__ucode_rom[2U] = 2U;
    // ...
    vlSelf->fsm__DOT__state = ((IData)(vlSelf->res)
   			? 0U : (3U & (vlSelf->fsm__DOT__ucode_rom
   				[(((IData)(vlSelf->fsm__DOT__state)
   						<< 1U)
   					| (IData)(vlSelf->d_in))]
   				>> 1U)));
    // ...
\end{lstlisting}

\noindent\textbf{Generalizing to HDL Designs.} Hardware descriptions can be written to target an application-specific technology library. A technology library is a series of definitions that map into existing hardware constructs called the \emph{technology}. When performing synthesis towards a technology library tools will map the described hardware to components in the library. The technology library may use FPGA-fabric specific constructs and
FPGA vendor's IP blocks written so that synthesis tool can translate
functionality directly to FPGA fabric. This may require changes to the HDL to meet fabric and design timing constrains.
When targeting ASICs, a hardware designer will explicitly write HDL to fully describe the intent of the hardware being described. This allows full control of the hardware that is generated by the synthesis tool.

When using technology-specific HDL styles, the higher level semantics of the language are lost. Instead, designers make conscious choices based on the microarchitectural nuances of the implementation. As high level semantics begin to be lost, so are identifiable constructs which are traditionally converted into coverage metrics when traditional fuzzing is performed. As a netlist is flattened into discrete components which describe a particular microarchitecture, lost are possible conversion paths that yield dedicated basic blocks and thus potential control-flow information. As diverging control-flow paths are flattened into one single linear construct, code coverage becomes a much less important metric for guiding a fuzzer. This motivates the introduction of a new metric.

\noindent\textbf{Value-Guided Fuzzing.} This paper presents a Value-Guided Fuzzing (\projectname{}) approach, which attempts to solve the mentioned issue of implicit control-flow not being captured by control-flow coverage guided  software fuzzers by reconstructing missing control-flow information via analysis of critical signals in the design. For the FSM above, \projectname{} looks at changes in the state registers to accomplish this goal and relies on the insight that once state register value changes can be tracked, one can infer traversal along a new path and, thus, reconstruct control-flow on an otherwise flattened design. 

\projectname{} allows us to dynamically rebuild an equivalent model of a control-flow graph (CFG) by tracking changes in critical signals under a stimulus. To acheive this \projectname{} requires an HDL model, a simulator, a test harness, and a slightly modified version of American Fuzzy Lop (AFL)~\cite{googleafl}. A note regarding the test harness: it is independent of the underlying hardware and only requires input and the clocks that drive those inputs, assertions, and information regarding the critical signals to track to populate the AFL hashmap. Internally, the test harness operates much like binary-only fuzzing by injecting callbacks on signals that are being tracked. The callback gets executed every time the signal changes value and this value is then used to track coverage. 

\noindent\textbf{Contributions.} In short, the contributions of this paper are:

\begin{itemize}
    \item A reexamination of the assumptions made by previous fuzzing approaches showing how these assumptions fail under application-specific HDL coding styles, and how these affect standard control-flow guided fuzzing.
    \item An improved coverage information metric to guide hardware fuzzers which provides better feedback for testcase generation.
    \item The introduction of \projectname{}, a value-guided simulator-independent fuzzing harness for hardware. We show that \projectname{} provides better coverage tracking than traditional control-flow based coverage metrics.
    \item The evaluation of \projectname{} in different design methodologies, showing its coverage mechanism quickly leads fuzzers to trigger the configured assertion.
\end{itemize}


%% file: vgf/vgf.tex
Current fuzzing mechanisms for hardware employ conversion of HDLs into C/C++ equivalent models. Although there are current limits to how accurate this conversion is, for the remainder of this work we assume that accurate conversion is an engineering issue and can be fixed. We also do not concern ourselves as to the speed/accuracy tradeoff of a simulator. However, because of the different ways HDL is written, the generated C/C++ code may not contain sufficient control-flow edges for a fuzzer which derives coverage information using control-flow graph edges to eliminate. By reexamining this conversion process, we arrive at the conclusion that a new metric for coverage is necessary.


\subsection{Conceptual Derivation}

Conceptual derivation of Value Guided Fuzzing (VGF) draws inspiration from an observation on how digital hardware operates. As time moves forward, different signals internal to the hardware change their value in response to a stimulus. The stimulus can be either clocked inputs for sequential designs or in the case of purely combinational logic drivers for asynchronous signals. VGF makes use of changes in signals to evaluate the effect inputs have on the design. For the remainder of this paper we call the set of values signals have on the design at a given time the \emph{state} of the design. Note that this \emph{state} does not necessarily reflect the state of a state machine, but the general state the hardware is in at a given point in time. Previous approaches attempt to capture state information by utilizing code coverage \cite{trippel2022fuzzing}, or pre-determined hardware events \cite{10.1145/3240765.3240842}. However, these metrics are not generalizable to designs that utilize flattened netlists and have limitations when dealing with multiple clock domains.

Generally speaking, consider an HDL core with signal $s \in \mathbb{S}$, where $\mathbb{S}$ is the set of signals. Let $\mathbb{K}_s$ be the set of possible states for signal $s$ with $k_s \in \mathbb{K}_s$ the current state of the signal. The current state of the hardware is then the given by $\cup_{s\in\mathbb{S}} k_s$. After a simulation timestep, function $f_s: \mathbb{K}_s \to \mathbb{K}_s$ is applied to every signal in the design. That is, on every timestep a signal is changed according to its driver logic. Our approach uses a set of signals $\mathbb{T} \subset \mathbb{S}$. Then, for every variation $f_s(k_i \in \mathbb{K}_s) \to k_{i+1} \in \mathbb{K}_s$ we compute a \emph{bucket} using a compression function $H: (s, k) \to \mathbb{N}_n$ where $s\in\mathbb{T}$, $k\in\mathbb{K}_s$, and $\mathbb{N}_n$ is the set of buckets representing a discovered transition, to provide coverage feedback to the AFL fuzzer.

Traditional software strategies use the execution of basic blocks as a measure of coverage. We can describe a program as a set of basic blocks $\mathbb{V}$ and edges connecting those basic blocks $\mathbb{E}$. Under AFL's fuzzing methodology an edge $e\in\mathbb{E}$ is discovered by giving unique identifiers to each basic block and mixing basic block identifiers into a compression function. An edge $e$ connecting basic blocks $b_1,b_2 \in \mathbb{B}$ is discovered by the application of a function $g: \mathbb{B}\times\mathbb{B} \to \mathbb{N}_n$. As previously discussed, when hardware is converted to software which generates basic blocks, the execution of a basic block implies a change in the hardware state. That is, new basic blocks are executed to set new signal states. Consequently, the operation $\mathbb{B} \times \mathbb{B}$ is equivalent to $\mathbb{T} \times \mathbb{K}_s$ from our approach, the latter of which evaluates to the transformation of the signal into a new state. Therefore, the methodology used by VGF is equivalent to that of traditional fuzzing while maintaining hardware semantics and still being able to discover new hardware states under the lack of proper basic block definitions.

Naively implementing this fuzzing strategy would require a different instrumentation pass on AFL, where desired signals are tracked for changes. Those changes would then be used to populate the hashmap that guides the fuzzer's mutational algorithm. Unfortunately, performing this type of instrumentation requires extra information which is not available at instrumentation time. HDL to C++ conversion tools, such as Verilator, would need to be modified to issue this instrumentation instead using hints from the hardware designer or tester. Fortunately, a solution can be architected without modifying existing simulation infrastructure. Moreover, as we will show, our implementation of VGF is independent of both the simulator employed \emph{and} the HDL chosen.

\subsection{Signal Selection Strategy}
\label{sec:svf}
\import{background/svf/}{svf.tex}

%% file: background/svf/svf.tex
A core concept of VGF is the signal selection to use to track state changes in the design under test. That is, relevant signals must be selected to make up $\mathbb{T}$.
One possible method is to use the knowledge of the design and choose $\mathbb{T}$ intuitively.
However, the complexity of hardware designs make it challenging to choose the 
relevant signals. 

We use the correctness/security property to automatically derive $\mathbb{T}$ 
using static dependency analysis.
Specifically, given a property expression, $\mathbb{P}$, we identify the set of signals, 
$\mathbb{PS}$, 
that appear in $\mathbb{P}$. 
Tracking the signals that impact the signals in $\mathbb{PS}$ can be 
used to guide the fuzzer towards generating inputs that can violate $\mathbb{P}$.
However, it is not desirable to use all the signals that may impact $\mathbb{PS}$ as 
this may yield high overhead. 
We, therefore, define a distance metric called {\em minimum distance to property signals} ($\texttt{PD}$) to quantify the property-relevance of a signal in $\mathbb{S} \setminus \mathbb{PS}$, defined as
\begin{equation}
\text{PD}^A(s,\mathbb{PS})=
\begin{cases}
0 &  s \in \mathbb{PS} \\
Min(Dis) & s \not\in \mathbb{PS},
\end{cases}
\end{equation}
where $A$ denotes the analysis type (data-flow or control-flow), $Dis=\{ d \ | \ \exists s' \in \mathbb{PS}. \ Reachable(G^A,s,s') \ \wedge \ d=MinDist(s,s') \}$, and $G^A$ denotes the dependency graph based on analysis $A$, which can be data-flow analysis, control-flow analysis, or a combination of the two.



Once $\texttt{PD}$ has been defined for each signal, it is necessary to establish a quantitative metric for selecting the appropriate set of property signals $\mathbb{T}$ in either data-flow or control-flow analysis. 
To achieve this, we introduce a threshold value $\tau$, which serves to filter out signals whose minimum distance to the property signals is less than or equal to a certain value.

%% file: prototype/prototype.tex
The following describes the architecture of our implementation of Value-Guided Fuzzing (VGF). We show how our VGF harness integrates into existing simulators and how reliability and integrity of the simulation is maintained during fuzzing. We present two different implementations of VGF: \vgfa{} and \vgfb{}.

\subsection{Binding Into a Simulator's Runtime}
HDL standards provide the means to extend language simulators with new functionality. The Verification Procedural Interface (VPI), Direct Programming Interface (DPI), and VHDL Procedural Interface (VHPI) and Foreign Language Interface (FLI) provide a mechanism for Verilog, SystemVerilog, and VHDL simulators to be extended, respectively. Since these application programming interfaces (APIs) provide different means to the same end, we will only introduce VPI.

\import{background/vpi/}{vpi.tex}

Although it is possible to implement VGF directly using VPI, we use cocotb \cite{rosser2018cocotb} as a middle layer to support as many simulation environments as possible. Depending on the simulation environment, cocotb chooses which interface to utilize to communicate with the simulator. This gives us the opportunity to seamlessly fuzz any HDL as long as there is simulator support for it on a simulator supported by cocotb.

\subsection{Overview of the \vgfa{} Harness}

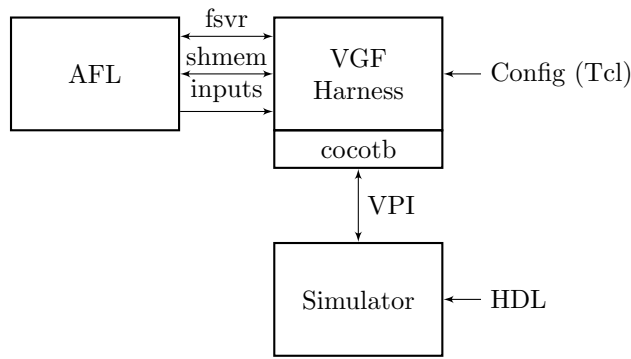
\begin{figure}[h]
    \centering
    \import{img/}{overview.tex}
    \caption{General \projectname{} structure. We use cocotb as an abstraction of the simulator API to connect to it while providing a harness to communicate with AFL.}
    \label{fig:vgf_overview}
\end{figure}

We show the overall structure of our fuzzing scheme in Figure \ref{fig:vgf_overview}. There are three components to our implementation: an HDL simulator, a fuzzer, and the VGF harness. The VGF harness is implemented on top of cocotb \cite{rosser2018cocotb}, which provides a simulator-agnostic layer to the VPI, as well as DPI for SystemVerilog simulators, VHPI and FLI for VHDL simulators. The harness uses cocotb to communicate with the simulator while providing an interface to the AFL fuzzer.


An external Tcl configuration script is used to configure the harness. The script defines clocks and reset signals, if any. The harness is able to handle both synchronous and asynchronous designs, as well as multiple clock domains. The configuration script also dictates the conditions the harness must check for, as well as the signals to be monitored for changes to populate AFL's control-flow hashmap.

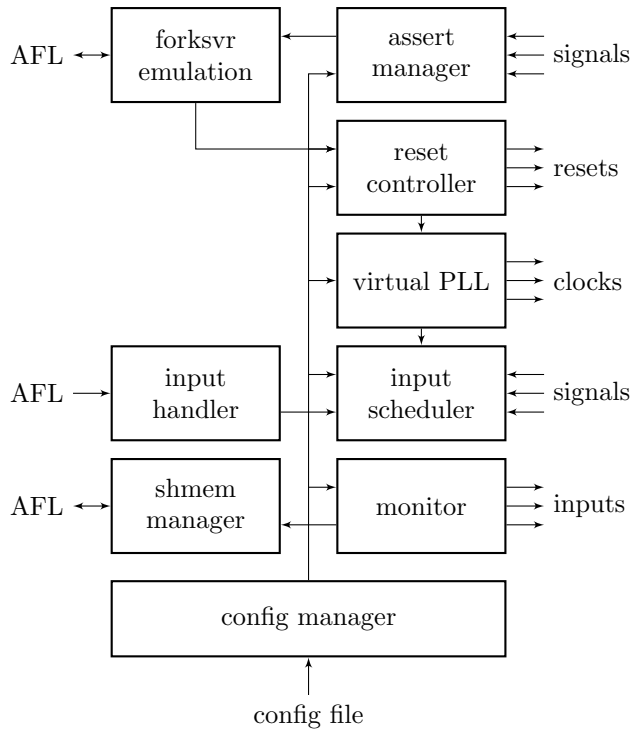
\begin{figure}[h]
    \centering
    \import{img/}{architecture.tex}
    \caption{The \projectname{} harness architecture.}
    \label{fig:vgf_architecture}
\end{figure}

The harness can operate in \emph{standalone mode}, or in \emph{guided fuzzing mode}. The former allows hardware designers and testers to manually provide inputs to the design. The latter allows integration with AFL for automated guided fuzzing. Figure \ref{fig:vgf_architecture} shows the detailed architecture of the harness. Once launched, the harness reads the environment variable containing the System V shared memory region (shmem) exported by AFL as its hashmap making it accessible, as well as providing a bridge, to the AFL fork server (fsvr) pipes. The following sections further detail portions of the harness.

\subsubsection{Fork Server and Resets}

We emulate the fork server protocol by performing reads and writes to the pipes shared with AFL. However, unlike traditional fuzzing, we do not utilize the fork server to spawn new instances of the simulator. Instead, the simulator process remains static in memory. Receiving a spawn command from AFL triggers a design reset. This is done through the use of a virtual reset controller. The configured reset signals are asserted, bringing the design to a known default state. Known default states are not tested by VGF, as these can be verified through standard functional testing. The virtual reset controller supports both active high and active low resets, which can be synchronous or asynchronous. Currently, when a reset is desired, all resets are held active for one period of the slowest clock on the design. Although currently not implemented in our harness, it is possible to use the reset controller to place the design in a custom initial configuration, such as initializing dynamically initializing the contents of a memory or a register file.

\subsubsection{Clocking and Input Scheduling}
Imperative to the fidelity of the simulation is the use of proper clocks. Our VGF harness creates a virtual phase locked loop (PLL) with its own internal reference clock. This virtual PLL object is used to create all clocks in the design, each with independent frequencies, duty cycles, and phase shifts as required by the configuration. The virtual PLL can further generate an optional \emph{clock stable} signal which can be used by the design under test to enable any clock dependent logic.

The VGF harness expects inputs to the design to be provided through regular standard input as per AFL's behavior. The harness contains an internal input scheduler which operates in a round-robin fashion. The configuration file for the harness defines all inputs to the design and constrains them to a clock source, defining input clock domains. For every input clock domain, if the domain can accept a new input, then the harness slices the data obtained from AFL into the total widths of the inputs and sends it. If there is not enough data on the input buffer, the data is padded by zeros. The input controller for the clock domain further slices the block of data and sends it to the inputs on that clock domain themselves. This form of scheduling and driving data into inputs produces deterministic results during testing. That is, simulation results are reproducible for any input provided to the design.

\subsubsection{Handling Designs with Differential Clocks}
A differential signal is one that uses complimentary voltage pairs. Differential signals are used to avoid noise in high speed serial and parallel transmission lines, as the noise gets cancelled when the differential signals are mixed on inputs. Some designs, specially those targeting ASICs use differential clock signals to drive latches. These clock signals are propagated through a \emph{clock network} internal to the design and used with latching logic to implement flip-flops. Our harness can be configured to use differential clocks by adding a clock signal to the inverted clock output which is phase shifted by $180^\circ$ with a duty cycle of $1 - h$, where $h$ is the duty cycle of the base clock in the pair.

\subsubsection{Handling Asynchronous Design}
Asynchronous designs are those that do not use any clock signals. Currently, our harness supports these designs by sending data into the inputs using the virtual PLL clock as a reference.
If a subset of the inputs on the design are asynchronous, then these are grouped into the first clock domain.

\subsubsection{Signal Monitor and AFL Feedback}
The configuration file for the harness specifies which signals to monitor for changes, as well as the \emph{weight} of those signals in the design. 
Selection of these signals using static analysis is explained in Section \ref{sec:svf}. When starting, a callback function is added to the signal change event for the specified signals. The callback function proceeds to use a unique signal identifier as well as the value of the signal as a means of indexing into the hashmap exported by AFL.

During the fuzzing process, the simulator fires an event every time one of the tracked signals changes. Our harness services that event incrementing a bucket of the hashmap. Some of the callback functions we implemented make use of the weight parameter to modify the value in the hashmap.

Lastly, the configuration file for the harness provides a series of conditions to check for in a format similar to that of SystemVerilog assertions. The harness monitors these conditions and reports to AFL if one is met. Hardware does not crash the same way software does. If a checked condition is encountered during the fuzzing process, the simulation \emph{does not} stop. Instead, the fuzzing harness creates a \emph{fake} signal error code which gets sent to AFL. AFL sees this as if it was a crash, creating a record containing the input which was sent to the harness that caused the fault. This input can be replayed at a later time with the harness in manual fuzzing mode.

\subsection{The \vgfb{} Harness}
An inherent limitation of interfacing with a simulator through their programming interface is the resulting speed of simulation. This can be detrimental if a fast fuzzing campaign is desired. To address speed concerns we introduce an alternate implementation which uses a modified Verilator output model which reports signal changes to an intermediate bridge which communicates with the fuzzing engine. The bridge in question allows for fast software emulation of hardware while providing the necessary fuzzer feedback. As we will demonstrate later, \emph{any Verilator-based harness will fail to capture bugs which manifest due to timing events such as in designs with multiple clock domains}. This is because Verilator achieves high-speed simulation capabilities at the expense of ignoring important portions of the SystemVerilog standard.

The \vgfb{} collects signal changes by checkpointing signals in the design before and after the execution of the event loop in Verilated code. The collected changes are asynchronously sent to the intermediate bridge without having to interrupt the simulation. Once the bridge collects all signal change data it sends the necessary feedback to the fuzzing engine. This achieves higher fuzzing throughput at the expense of accuracy in simulation.

\subsection{Detecting Property Relevant Signals}
\label{sec:svfanalysis}

\input{background/svf/svfImpl}

%% file: background/vpi/vpi.tex
The Verification Procedural Interface (VPI) is a standard defined as part of
IEEE Standard 1800 \cite{ieee:1800}. Previously known as the Verilog
Programmable Interface, VPI provides an API
which allows direct interfacing with Verilog and SystemVerilog simulators.
Although specific to the Verilog family of languages, VPI is also available on
VHDL simulators making the interface widely available. VPI is available
in commercial simulators and open-source simulators such as Icarus
Verilog. At the time of writing, Verilator has limited support for VPI.



%% file: img/overview.tex
\begin{tikzpicture}[
		component/.style = {
			draw,
			rectangle,
			thick,
			minimum height=1.5cm,
			text width=2cm,
			text centered
		}
	]

	\node[style=component] (afl) at(-0.5, 0) {AFL};

	\node[style=component] (hrn) at(3, 0) {\projectname{} Harness};
	\node[style=component, minimum height=0.5cm] (cocotb) at(3, -1) {cocotb};

	\node[style=component] (sim) at(3, -3) {Simulator};
	
	\draw[latex'-] (sim.east) -- ++(0.5, 0) node[anchor=west] {HDL};
	\draw[latex'-] (hrn.east) -- ++(0.5, 0) node[anchor=west]
		{Config (Tcl)};

	\draw[latex'-latex'] (cocotb) -- node[anchor=west] {VPI} (sim);

	\def\comm {
		fsvr/latex'-latex',
		shmem/latex'-latex',
		inputs/-latex'%
	}

	\foreach \lbl\arrowtype [count=\y from 0] in \comm {
		\draw[\arrowtype]
				($ (afl.east) + (0, -0.5*\y+0.5) $)
				-- node[anchor=south] {\lbl}
				($ (hrn.west) + (0, -0.5*\y+0.5) $);
	}

\end{tikzpicture}

%% file: img/architecture.tex
\begin{tikzpicture}[
        component/.style = {
			draw,
			rectangle,
			thick,
			minimum height=1.25cm,
			text width=2cm,
			text centered
		},
	]
	
	\def\components{
	    assert manager,
	    reset controller,
	    virtual PLL,
	    input scheduler,
	    monitor%
	}
	
	\foreach \lbl [count=\y] in \components {
	    \node[style=component] (c\y) at(0, -1.5*\y) {\lbl};
	}
	
	\def\components{
	    forksvr emulation/c1,
	    shmem manager/c5,
	    input handler/c4%
	}
	
	\foreach \lbl\npos [count=\y from 6] in \components {
	    \node [style=component, left of=\npos, node distance=3cm]
	        (c\y) {\lbl};
	}
	
	\draw let \p1=(c5.center) in let \p2=(c6.center) in
	    node[style=component, text width=5cm, minimum height=1cm]
	    (c9)
	    at (0.5*\x1 + 0.5*\x2, \y1 - 1.5cm)
	    {config manager};
	
	\draw[-latex'] (c9) |- ($(c1.west) + (0, -0.25)$);
	\foreach \cnode in {c2.west,c4.west,c5.west} {
	    \draw[-latex'] let \p1=(c9.center) in
	        let \p2=(\cnode) in
	       (\x1, \y2 + 0.25cm) -- (\x2, \y2 + 0.25cm);
	}
	
	\draw[-latex'] let \p1=(c9.center) in
	    let \p2=(c2.west) in
	    (\x1, \y2 - 0.25cm) -- (\x2, \y2 - 0.25cm);
	\draw[-latex'] let \p1=(c9.center) in
	    let \p2=(c3.west) in
	    (\x1, \y2) -- (\x2, \y2);
	
	\draw[-latex'] ( $(c8.east) - (0, 0.25cm)$ )
	    -- ( $(c4.west) - (0, 0.25cm)$ );
	\draw[latex'-] ( $(c7.east) - (0, 0.25cm)$ )
	    -- ( $(c5.west) - (0, 0.25cm)$ );
	
	\draw[-latex'] (c6) |- ($(c2.west) + (0, 0.25)$);
	\draw[latex'-] ($(c6.east) + (0, 0.25) $) -- ($(c1.west) + (0,0.25)$);
	
	\draw[-latex'] (c3) -- (c4);
	\draw[-latex'] (c2) -- (c3);

    \def\simulatorside {
        latex'-/c1.east/signals,
        -latex'/c2.east/resets,
        -latex'/c3.east/clocks,
        latex'-/c4.east/signals,
        -latex'/c5.east/inputs%
    }
    
    \foreach \arrowstyle\nodecoord\lbl in \simulatorside {
        \draw[\arrowstyle] (\nodecoord) -- +(0.5, 0) node[anchor=west] {\lbl};
        \draw[\arrowstyle] let \p1=(\nodecoord) in
            (\x1, \y1 + 0.25cm) -- +(0.5cm, 0);
        \draw[\arrowstyle] let \p1=(\nodecoord) in
            (\x1, \y1 - 0.25cm) -- +(0.5cm, 0);
    }

    \def\aflside {
        latex'-latex'/c6.west/AFL,
        latex'-latex'/c7.west/AFL,
        latex'-/c8.west/AFL%
    }
    
    \foreach \arrowstyle\nodecoord\lbl in \aflside {
        \draw[\arrowstyle] (\nodecoord) -- +(-0.5, 0)
            node[anchor=east] {\lbl};
    }
    
    \draw[latex'-] (c9.south) -- +(0, -0.5)
        node[anchor=north] {config file};
    
\end{tikzpicture}

%% file: background/svf/svfImpl.tex
For the purposes of testing, we employed two different approaches to pick signals. First, we used a manual inspection approach where relevant signals for the tracking group were extracted. Then we cross-referenced our findings by using static analysis to extract and give weight to relevant signals.

\begin{lstlisting}[style=vcode, caption={A fragment of the AES128-T100 Design},label={lst:aes-t100}]
module aes_128(clk, reset, state, key, out);
    input          clk, reset;
    input  [127:0] state, key;
    output [127:0] out;
    reg    [127:0] s0, k0, out_reg;
    wire   [127:0] s1, s2, s3, s4, s5, s6, s7, s8, s9,
                   k1, k2, k3, k4, k5, k6, k7, k8, k9,
                   k0b, k1b, k2b, k3b, k4b, k5b, k6b, 
                   k7b, k8b, k9b, core_out;
    assign out = out_reg;
	always @(posedge clk)
		if(reset) out_reg <= {128{1'b0}};
		else if((k1[12:8] & k4[122:118]) == 5'b11111) 
			out_reg <= {128{1'b1}};  /* trojan payload */
		else if(~(out_reg ^ {128{1'b1}}))
			out_reg <= core_out;
    always @ (posedge clk) begin
        s0 <= state ^ key; k0 <= key;
    end
    expand_key_128
       a1 (clk, reset, k0, k1, k0b, 8'h1), /* ... */
       a10 (clk, reset, k9,   , k9b, 8'h36);
    one_round
       r1 (clk, reset, s0, k0b, s1), /* ... */
       r9 (clk, reset, s8, k8b, s9);
    final_round rf (clk, reset, s9, k9b, core_out);
endmodule
\end{lstlisting}

\subsubsection{Intuitive Signal Analysis}
The goal of the intuitive signal analysis, VGF-intuitive, 
is to show the performance of VGF when a user selects signals to track based on their own understanding of a design without the assistance of an external tool. For these experiments, each design is analyzed manually and a signal configuration is selected based on the perceived importance of each signal as it relates to the assertion a testbench checks to validate a design's proper execution.

As an example of how intuitive signal selection occurs, consider the following example based on the \texttt{AES128-T100} testbench, Listing~\ref{lst:aes-t100}. If we assume our assertion verifies the correct output of the AES-128 encryptions, then we can track the signals related to the relevant assertion, which is the output signal \lstinline[style=vcode]{out}. This signal is a function of the AES algorithm which is purely driven by the plaintext and encryption key. Therefore, the intuitive signal configuration to select would be based off of either the encryption key input, or the plaintext input. In this case, signal \lstinline[style=vcode]{k0}, representing a register holding the inputted key value, is selected for the VGF-intuitive test. 

\subsubsection{Design Static Analysis}

We use a static analysis tool, SVF \cite{svf}, to perform data-flow analysis and control-flow analysis, which provide the signal-relevant dependency information for the signal(s) of interest. We will refer to this method as VGF-static\_analysis.
Specifically, we use Verilator\footnote{Although Verilator generated code may not always be conducive to fuzzing that is guided by edge-based code coverage, it preserves the data-flow and control flow dependencies among signals as long as they are not optimized out.} to generate a software representation of the hardware and then use the clang compiler to generate its representation in LLVM IR. Static Value-Flow (SVF) is implemented on the top of LLVM and allows value-flow construction and pointer analysis to be performed in an iterative manner.  
SVF also enables scalable and precise inter-procedural analysis by leveraging sparse analysis, which provides a promising solution for analyzing large programs. 

\emph{Data-flow Analysis.}
SVF generates a sparse data-flow dependency graph, SVFG, based on the LLVM IR, which provides the def-use chains among the instructions while taking into account conservative points-to information.
Since Verilator generates a software representation using a {\tt struct} data type, it is easy to identify the specific hardware signals in the SVFG by matching the index that corresponds to the specific field representing the signal in the struct type with the index in the {\tt GetElementPtr} instructions.
Next, on the SVFG, we find all the paths starting from the {\tt load} instructions of the input signals to the {\tt store} instructions accessing other signals and compute the data-flow dependency relevant minimum distance to property, $PD$, where the graph $G^{DF}$ is the SVFG.

Consider the simple hardware design depicted in Listing \ref{lst:static_analysis_eg}, and suppose that data-flow analysis commences from output port {\tt out}. We observe that the signals {\tt tmp\_1}-{\tt tmp\_4} exhibit a data-flow dependency with respect to {\tt out}, and that their respective shortest data-flow paths, as extracted from the SVFG, are tmp\_1$\rightarrow$tmp\_4$\rightarrow$out, tmp\_2$\rightarrow$tmp\_3$\rightarrow$tmp\_4$\rightarrow$out, tmp\_3$\rightarrow$tmp\_4$\rightarrow$out, and tmp\_4$\rightarrow$out. These paths allow us to determine the $PD$ for {\tt tmp\_1}-{\tt tmp\_4} to be 2, 3, 2, and 1, respectively.
In contrast, the signal {\tt tmp\_5} does not exhibit any data-flow dependency with respect to {\tt out}, since there exists no data-flow path from {\tt tmp\_5} to {\tt out}.


\begin{lstlisting}[style=vcode, caption={A simple combinational design in Verilog to demonstrate data-flow dependency.},label={lst:static_analysis_eg}]
module comb(input din_0, din_1, output reg out);
  reg tmp_1, tmp_2, tmp_3, tmp_4, tmp_5;
  always @ (*) begin
    tmp_1 = din_0 & din_1;
	tmp_2 = din_0 | din_1;	
	tmp_3 = tmp_1 ^ tmp_2;
	tmp_4 = tmp_3 | tmp_1;
	tmp_5 = tmp_1 & din_1;
	out = tmp_4 ^ 1'b1;
  end
endmodule
\end{lstlisting}


To achieve property-directed value-guided fuzzing, we identify property relevant 
LLVM IR instructions by identifying the {\tt store} instructions that define signals 
that get used in an {\tt assert} expression. Then we find the weighted data-flow 
dependency data in the form of $(S_i,\texttt{weight}_{ij},S_j)$, where signal $S_i$ flows into 
signal $S_j$ and $\texttt{weight}_{ij}=1/(\texttt{PD}+1)$. 
The weight of a signal is then used by the harness to increment the corresponding bucket 
in the AFL control-flow hashmap in the amount of the weight when a change in the signal 
is detected.

\emph{Control-Flow Analysis.}
SVF is also capable of producing an Interprocedural Control-Flow Graph (ICFG) from LLVM IR, which can provide valuable information about a program's control-flow dependencies. 
Unlike data-flow analysis, which typically starts with identifying the {\tt GetElementPtr} instruction, the process for control-flow analysis involves identifying interesting {\tt Store} instructions with internal signal flows 
and creating {(\tt Store, GetElementPtr)} pairs. 
From there, all branch conditions in the predecessor basic blocks of these Store instructions are iteratively identified. 
Finally, operands for each branch condition are searched until a {\tt GetElementPtr} instruction representing the input variable is found. 
So, the graph, $G^{CF}$, for control-flow analysis, is a combination of the SVFG and the ICFG.

As the simple sequential hardware design in Listing \ref{lst:cfa_eg} shows, we assume that control-flow analysis starts from output {\tt out}, i.e., {\tt out} is in $\mathbb{PS}$. 
The signals {\tt state,tmp\_1,tmp\_2} have control-flow dependency from {\tt out}, and their corresponding shortest control-flow path, as extracted from ICFG, are state$\rightarrow$out, tmp\_1$\rightarrow$state$\rightarrow$out, and tmp\_2$\rightarrow$state$\rightarrow$out. 
These paths also determine the $PD$ for signals {\tt state,tmp\_1,tmp\_2} to be 1, 2, and 2, respectively.
However, the signal {\tt tmp\_3} has no control-flow dependency but has data-flow dependency with respect to {\tt out} since there is a data-flow but no control-flow path from {\tt tmp\_3} to {\tt out}.



\begin{lstlisting}[style=vcode, caption={A simple sequential design in Verilog to demonstrate control-flow dependency.},label={lst:cfa_eg}]
module seq(input clk, din_0, din_1, output out);
  wire tmp_1, tmp_2, tmp_3;
  reg [1:0] state;
  assign tmp_1 = din_0 && din_1;
  assign tmp_2 = din_0 || din_1;
  assign tmp_3 = tmp_1 && tmp_2;
  always @(posedge clk)
    if (din_0)
      if (tmp_1 && tmp_2)   state <= 2'b11;
      else                  state <= 2'b10;
    else if (din_1)         state <= 2'b01;
    else                    state <= 2'b00;
  always @(*)
    case(state)
      2'b00: out = tmp_3;
      /* ... */
    endcase
endmodule
\end{lstlisting}

Similar to the approach used in data-flow analysis, we then find the weighted control-flow dependency data in the form of $(S_i,weight_{ij},S_j)$, where signal $S_i$ flows into signal $S_j$  and $\texttt{weight}_{ij}=1/(\texttt{PD}+1)$ to achieve property-directed value-guided fuzzing according to control-flow dependency.

%% file: results/results.tex

In this section, we describe the experimental results of our methodology. We collected a series of designs with assertion checks on them and tested how many input mutations AFL needed to trigger the assertion. A description of the designs is shown in Table \ref{tb:test_designs} alongside their source.

\begin{table*}
    \centering
    \begin{tabularx}{\textwidth}{lXc}
    \toprule
    \textbf{Design} & \textbf{Description} & \textbf{Source} \\
    \midrule
     AES128-T100 & 128-bit AES encryption algorithm, Trojan causes ciphertext to be stuck at '1' once activated & \cite{TrustHub} \\
     lock\_case & FSM implementation using case statements, design gets unlocked with a unique input sequence & \cite{trippel2022fuzzing} \\
     lock\_micro & FSM implementation using microcode, design gets unlocked a unique input sequence  & VGF \\
     DAIO & Full duplex, fully symmetrical interface chip for the RX and TX of digital audio signals & \cite{v2c2016}  \\
     Dekker & Faulty algorithm for mutual exclusion of two processes
     & \cite{v2c2016} \\
     Unidec & Nondeterministic checker for unique decipherability & \cite{v2c2016} \\
     AES128-T2500 & 128-bit AES encryption, Trojan flips ciphertext's least significant bit after 15 clock cycles & \cite{TrustHub}\\
     RS232-T600 & Trojan activates with unique input sequence, sticks output signal at high and corrupts TX output & \cite{TrustHub} \\ 
     AE18\_core & AE18 8-bit Microprocessor, counter-based Trojan, when active flips output from write back & \cite{OpenCores}\\ 
     async\_fifo & A multi-clock domain FIFO. Head pointer overtakes tail if clocks have different frequencies & VGF \\
    \bottomrule
    \end{tabularx}
    \caption{Test designs used in this paper and their description.}
    \label{tb:test_designs}
\end{table*}

\subsection{Initial Experimental Run}
Results for the intial experimental run
with the intuitive configuration, VGF-intuitive, and static analysis configuration, VGF-static\_analysis using \vgfa{}
are shown in Table \ref{tb:exp_results}. In Table \ref{tb:exp_results_time}, the results for \vgfb{} are shown.
Testing was done on a 96 core server using Icarus Verilog \cite{williams2020icarus} and VCS \cite{synopsys2022} to evaluate our fuzzing methodology for \vgfa{} and Verilator to evaluate our methodology for \vgfb{}. 

\subsubsection{Performance of \vgfa{}-intuitive}



Selecting signals based on the user's intuition of which signals are relevant for the \vgfa{} configuration displays increased performance against {HWFuzz}. This result shows that even with a configuration of signals that are not necessarily optimal, \vgfa{} can, in most cases, still fuzz the target design with a high level of efficiency. We can conclude that considering the relevance of signals to a design's control-flow with respect to the property being tested is an important aspect of fuzzing hardware. \vgfa{} is able to more quickly identify testcases capable of triggering the assertion by limiting itself to tracking the state of these signals as a change to the relevent signals. As a result, this will affect the property the assertion is checking more than other signals.

\subsubsection{Performance of \vgfa{}-static\_analysis}

This VGF configuration leverages signal dependency information obtained from static analysis, leading to a substantial improvement in performance. 
 
Specifically, by using precise control- and data-flow dependency information, {\vgfa{}-static\_analysis} selects an appropriate set of signals and assigns appropriate weights to each signal, enabling it to more effectively monitor changes in the signals within the design. 
As a result, {\vgfa{}-static\_analysis} requires fewer execution cycles to reveal faults inside the designs compared with {HW-Fuzz}.
Thus, VGF acheives superior performance by leveraging precise information about signal dependencies and weights obtained from static analysis.


\subsubsection{Comparing Performances of \vgfa{}-static\_analysis to \vgfa{}-intuitive}
Overall, signals obtained through a static analysis pass improves the performance of \vgfa{}.
This difference in performance can be attributed to the fact that the signal set with corresponding weights selected by \vgfa{}-intuitive is a subset of the signal set selected by \vgfa{}-static\_analysis. With the intuitive approach, only a limited number of relevant signals were selected. As a result, the VGF-intuitive approach may leave out, or not give enough importance to, crucial signals that are relevant to the control-flow of the design which results in a worse performance. In fact, \vgfa{}-intuitive provides similar performance to \vgfa{}-static\_analysis with a small value for the threshold $\tau$. Additionally, both methods of \vgfa{} prove able to handle benchmark \texttt{async\_fifo} whereas HW-Fuzz cannot. HW-Fuzz relies on Verilator for hardware simulation which is unable to handle multiple asynchronous clock domains. By leveraging Icarus Verilog as a simulator, \vgfa{} is able to handle benchmarks which HW-Fuzz cannot.

\begin{table}[htb]
    \begin{center}
    \begin{threeparttable}
        \begin{tabularx}{\columnwidth}{Xccc}
        \toprule

        \textbf{Design} & \textbf{HWFuzz} & \textbf{\vgfa{}-intuitive} & \textbf{\vgfa{}-static} \\
         \midrule
         AES128-T100 & 12158 & 20263  & 2101$^*$ \\
         lock\_case & 10137 & 8711  & 3213$^\dagger$ \\
         lock\_micro & 10704 & 5185 & 3167$^\dagger$ \\ 
         DAIO & 5380 & 1415 & 718$^*$ \\
         Dekker & 23 & 49 & 56$^*$\\
         Unidec & --  & -- & -- \\
         AES128-T2500 & 14702 & 1331 & 57$^\ddagger$\\
         RS232-T600 & 305299 & 92 & 53$^\dagger$ \\
         AE18\_core & 2065 & 91 & 64$^\dagger$\\
         async\_fifo & - & 4703 & 3065 \\
         \bottomrule
        \end{tabularx}
    \begin{tablenotes}
        \item[$*$] best result from data-flow analysis configuration
        \item[$\ddagger$] best result from control-flow analysis configuration
        \item[$\ddagger$] best result from combined data- and control-flow analysis configuration
    \end{tablenotes}
    \end{threeparttable}
    \caption{Average number of execution cycles to trigger fault in design from \vgfa{} intuitive configuration, HWFuzz, and \vgfa{} static\_analysis configuration. Average of 5 test runs per design and approach. Unidec and async\_fifo benchmarks were timed out after 24 hours. For all experiments $\tau=max/8$. 
    }
    \label{tb:exp_results}
    \end{center}
\end{table}

\begin{table}[htb]
    \begin{center}
    \begin{threeparttable}
        \begin{tabularx}{\columnwidth}{X*{6}{c}}
        \toprule
        \multirow{2}{*}{\textbf{Design}} &
        \multicolumn{2}{c}{\textbf{HWFuzz}} & 
        \multicolumn{2}{c}{\textbf{\vgfa{} / \vgfb{}}} \\
         &
         \textbf{Execs} & 
         \textbf{Time} &
         \textbf{Execs} & 
         \textbf{Time}  \\
         \midrule
         AES128-T100 & 12158 & 19.31 & 2101 / -- & 5.2 / -- \\
         lock-case & 10137 & 12.31 & 3213 / 9075 & 89.4 / 9.44  \\
         lock-micro & 10704  & 12.25 & 3167 / 4799 & 60.4 / 6.56 \\
         DAIO & 5380 & 9.78  & 718 / 828 & 11.8 / 5.17 \\
         Dekker & 23 & 4.20 & 56 / 105 & 5.2 / 4.53 \\
         Unidec & --  & -- & -- / -- & -- / -- \\ 
         AES128-T2500 & 14702 &  30.06 & 57 / 6858 & 9.6 / 9.99 \\
         RS232-T600 & 305299 &  267.04  & 53 / 133 & 5.0 / 4.52 \\
         AE18-core & 2063 & 2.58 & 64 / 217 & 5.0 / 4.51 \\
         async\_fifo & -- & -- & 3065 / -- & 68.6 / -- \\
         \bottomrule
        \end{tabularx}
    \end{threeparttable}
    \caption{Average number of execution cycles and seconds required to trigger a fault for \vgfa{}, \vgfb{} and HWFuzz. Average of 5 test runs per design and approach. Benchmarks were timed out after 24 hours. For all experiments $\tau=max/8$.} 
    \label{tb:exp_results_time}
    \end{center}
    \vspace{-0.7cm}
\end{table}


\begin{table*}
    \centering
    \begin{tabularx}{\textwidth}{lXc}
    \toprule
    \textbf{Signal(s) Tracked} & \textbf{Description} & \textbf{Execution Cycles} \\
    \midrule
    aes\_128.k0 & Latch that holds the value of the inputted encryption key & 20263 \\
    aes\_128.a4.v0 & Intermediate value set during calculation of the fourth round key expansion & 47030 \\
    aes\_128.a1.v0 & Intermediate value set during calculation of the first round key expansion& 55977 \\
   \multicolumn{1}{p{3.25cm}}{aes\_128.a1.v0 \&\newline  \hspace*{0.5cm} aes\_128.a4.v0 } & The same intermediate value from both the first and fourth around key expansion tracked together & 97356 \\
    aes\_128.a3.S4\_0.out & Output of the byte substitution step in the third round key exansion & Timeout \\
    \midrule[0.1pt]
    \multicolumn{1}{p{3.25cm}}{aes\_128.k0 \&\newline \hspace*{0.5cm} aes\_128.k9 \&\newline \hspace*{0.5cm} aes\_128.s0} & Initial key, ninth round key, and initial plaintext tracked together (selected by static analysis) & 2101 \\
    \bottomrule
    \end{tabularx}
    \caption{The effect of tracked signal on average execution time for the \lstinline[style=vcode]{AES128-T100} benchmark. Averaged over 5 iterations.}
    \label{tb:exp_results_signal}
\end{table*}

\begin{figure*}[bt]
\begin{center}
\begin{tikzpicture} 
  \begin{axis}[      
        ybar=0.5,
        ymin=1, ymax=300,
        ylabel={\# selected signals},
        ymode=log,
        x tick label style={rotate=15,anchor=east},
        x label style={at={(axis description cs:0.5,-0.2)},anchor=north},
        width=\textwidth,      
        height=1.65in,      
        symbolic x coords={{AES128-T100, DAIO, Dekker, Unidec, AES128-T2500, RS232-T600, AE18\_core, asyn\_fifo}}, 
        legend style={
            at={(0.5,1)},
            anchor=north east,
            legend columns=3,
            nodes={
                scale=0.8,
                transform shape
            }
        },
        xtick=data,
        enlarge y limits = 0,
        ]
        \addplot+[pattern=north east lines, pattern color=.] coordinates {(AES128-T100,243) (DAIO,8) (Dekker,3) (Unidec,3) (AES128-T2500,138) (RS232-T600,27) (AE18\_core,171) (asyn\_fifo, 9)};
        \addplot+[pattern=north west lines, pattern color=.] coordinates {(AES128-T100,147) (DAIO,6) (Dekker,3) (Unidec,3) (AES128-T2500,112) (RS232-T600,27) (AE18\_core,161) (asyn\_fifo, 7)};
        \addplot+[pattern=crosshatch, pattern color=.] coordinates {(AES128-T100,7) (DAIO,5) (Dekker,2) (Unidec,3) (AES128-T2500,11) (RS232-T600,12) (AE18\_core,10) (asyn\_fifo, 4)};
        \addplot+[pattern=dots, pattern color=.] coordinates {(AES128-T100,3) (DAIO,4) (Dekker,1) (Unidec,1) (AES128-T2500,4) (RS232-T600,2) (AE18\_core,2) (asyn\_fifo, 4)};
        \addplot+[pattern=crosshatch dots, pattern color=.] coordinates {(AES128-T100,1) (DAIO,2) (Dekker,1) (Unidec,1) (AES128-T2500,1) (RS232-T600,2) (AE18\_core,2) (asyn\_fifo, 3)};
        \legend{$\tau=max$,$\tau=max/2$,$\tau=max/4$,$\tau=max/8$, $\tau=min$}
  \end{axis}
\end{tikzpicture}
\caption{The number of selected signals with respect to diffetent threshold values $\tau$ for data-flow analysis configuration. Designs \textbf{lock\_case} and \textbf{lock\_micro} are not listed since there is no data-flow dependent signals for the target signal.  Y-axis is logarithmic (minimum \# of signals is one).}
\label{fig:numsignals}
\end{center}
\end{figure*}
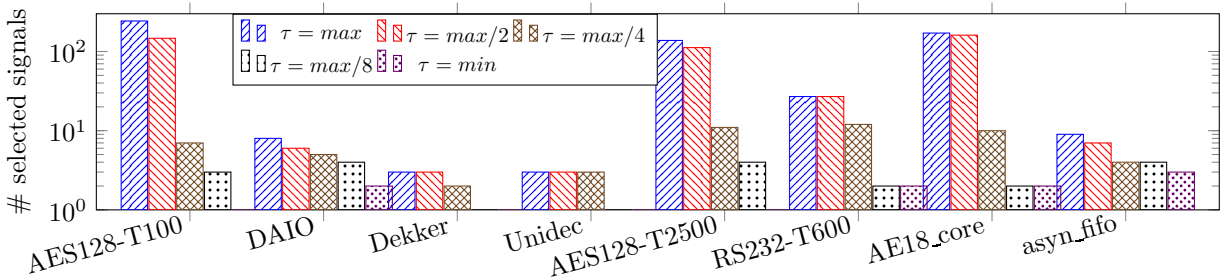


\begin{figure*}[ht]
\centering
\begin{subfigure}[t]{0.325\textwidth}
\begin{tikzpicture} 
  \begin{axis}[      
        ybar=0.5,
        ymin=0, ymax=200000,
        ylabel={\# execs},
        ymode=log,
        title={AES128-T100}, 
        x tick label style={rotate=45,anchor=east},
        width=\columnwidth,
        height=1.65in,      
        symbolic x coords={DFA, CFA, DCFA},
        legend style={
            at={(1,0)},
            anchor=south east,
            nodes={
                scale=0.6,
                transform shape
            }
        },
        xtick=data,
        enlargelimits=0.25,
        ]
    \addplot+[pattern=north east lines, pattern color=.] coordinates {(DFA,45243) (CFA,124546) (DCFA,146846)};
    \addplot+[pattern=north west lines, pattern color=.] coordinates {(DFA,3438) (CFA,156284) (DCFA,146846)};
    \addplot+[pattern=crosshatch, pattern color=.] coordinates {(DFA,53030) (CFA,129606) (DCFA,146846)};
    \legend{$\tau=max/4$,$\tau=max/8$, $\tau=min$}
  \end{axis}
\end{tikzpicture}
\end{subfigure}%
\hfill%
\begin{subfigure}[t]{0.325\textwidth}
\begin{tikzpicture} 
  \begin{axis}[      
        ybar=0.5,
        ymin=0, ymax=50000,
        ylabel={\# execs},
        ymode=log,
        title={lock\_case}, 
        x tick label style={rotate=45,anchor=east},
        x label style={at={(axis description cs:0.5,-0.2)},anchor=north},
        width=\columnwidth,      
        height=1.65in,      
        symbolic x coords={DFA, CFA, DCFA},
        legend style={
            at={(1,0)},
            anchor=south east,
            nodes={
                scale=0.6,
                transform shape
            }
        },
        xtick=data,
        enlarge x limits=0.25,
        enlarge y limits=0,
        ]
    \addplot+[pattern=north east lines, pattern color=.] coordinates {(DFA,1) (CFA,10408) (DCFA,10408)};
    \addplot+[pattern=north west lines, pattern color=.] coordinates {(DFA,1) (CFA,10408) (DCFA,10408)};
    \addplot+[pattern=crosshatch, pattern color=.] coordinates {(DFA,1) (CFA,10408) (DCFA,10408)};
    \legend{$\tau=max/4$,$\tau=max/8$, $\tau=min$}
  \end{axis}
\end{tikzpicture}
\end{subfigure}%
\hfill%
\begin{subfigure}[t]{0.325\textwidth}
\begin{tikzpicture} 
  \begin{axis}[      
        ybar=0.5,
        ymin=0, ymax=50000,
        ylabel={\# execs},
        ymode=log,
        title={lock\_micro}, 
        x tick label style={rotate=45,anchor=east},
        width=\columnwidth,      
        height=1.65in,      
        symbolic x coords={DFA, CFA, DCFA}, 
        legend style={
            at={(1,0)},
            anchor=south east,
            nodes={
                scale=0.6,
                transform shape
            }
        },
        xtick=data,
        enlarge x limits=0.25,
        enlarge y limits=0,
        ]
    \addplot+[pattern=north east lines, pattern color=.] coordinates {(DFA,1) (CFA,9908) (DCFA,9908)};
    \addplot+[pattern=north west lines, pattern color=.] coordinates {(DFA,1) (CFA,9908) (DCFA,9908)};
    \addplot+[pattern=crosshatch, pattern color=.] coordinates {(DFA,1) (CFA,9908) (DCFA,9908)};
    \legend{$\tau=max/4$,$\tau=max/8$, $\tau=min$}
  \end{axis}
\end{tikzpicture}
\end{subfigure}
\\
\begin{subfigure}[t]{0.325\textwidth}
\begin{tikzpicture} 
  \begin{axis}[      
        ybar=0.5,
        ymin=0, ymax=3000,
        ylabel={\# execs},
        ymode=log,
        title={DAIO}, 
        x tick label style={rotate=45,anchor=east},
        x label style={at={(axis description cs:0.5,-0.2)},anchor=north},
        width=\columnwidth,      
        height=1.65in,      
        symbolic x coords={DFA, CFA, DCFA}, 
        legend style={
            at={(0,1)},
            anchor=north west,
            nodes={
                scale=0.6,
                transform shape
            }
        },
        xtick=data,
        enlargelimits=0.25,
        ]
    \addplot+[pattern=north east lines, pattern color=.] coordinates {(DFA,718) (CFA,1969) (DCFA,2362)};
    \addplot+[pattern=north west lines, pattern color=.] coordinates {(DFA,718) (CFA,1362) (DCFA,2362)};
    \addplot+[pattern=crosshatch, pattern color=.] coordinates {(DFA,987) (CFA,1142) (DCFA,2362)};
    \legend{$\tau=max/4$,$\tau=max/8$, $\tau=min$}
  \end{axis}
\end{tikzpicture}
\end{subfigure}%
\hfill%
\begin{subfigure}[t]{0.325\textwidth}
\begin{tikzpicture} 
  \begin{axis}[      
        ybar=0.5, 
        ylabel={\# execs},
        ymode=log,
        title={Dekker}, 
        x tick label style={rotate=45,anchor=east},
        x label style={at={(axis description cs:0.5,-0.2)},anchor=north},
        width=\columnwidth,
        height=1.65in,      
        symbolic x coords={DFA, CFA, DCFA}, 
        legend style={
            at={(1,1)},
            anchor=north east,
            nodes={
                scale=0.6,
                transform shape
            }
        },
        xtick=data,
        enlargelimits=0.25,
        ]
    \addplot+[pattern=north east lines, pattern color=.] coordinates {(DFA,62) (CFA,70) (DCFA,60)};
    \addplot+[pattern=north west lines, pattern color=.] coordinates {(DFA,60) (CFA,69) (DCFA,60)};
    \addplot+[pattern=crosshatch, pattern color=.] coordinates {(DFA,60) (CFA,60) (DCFA,60)};
    \legend{$\tau=max/4$,$\tau=max/8$, $\tau=min$}
  \end{axis}
\end{tikzpicture}
\end{subfigure}%
\hfill%
%
\begin{subfigure}[t]{0.325\textwidth}
\begin{tikzpicture} 
  \begin{axis}[      
        ybar=0.5,
        ylabel={\# execs},
        ymode=log,
        title={AES128-T2500}, 
        x tick label style={rotate=45,anchor=east},
        x label style={at={(axis description cs:0.5,-0.2)},anchor=north},
        width=\columnwidth,
        height=1.65in,      
        symbolic x coords={DFA, CFA, DCFA}, 
        legend style={
            at={(1,1)},
            anchor=north east,
            nodes={
                scale=0.6,
                transform shape
            }
        },
        xtick=data,    
        enlargelimits=0.25,
        ]
    \addplot+[pattern=north east lines, pattern color=.] coordinates {(DFA,190) (CFA,484) (DCFA,57)};
    \addplot+[pattern=north west lines, pattern color=.] coordinates {(DFA,154) (CFA,412) (DCFA,57)};
    \addplot+[pattern=crosshatch, pattern color=.] coordinates {(DFA,47) (CFA,62) (DCFA,57)};
    \legend{$\tau=max/4$,$\tau=max/8$, $\tau=min$}
  \end{axis}
\end{tikzpicture}
\end{subfigure}%
\\
\begin{subfigure}[t]{0.325\textwidth}
\begin{tikzpicture} 
  \begin{axis}[      
        ybar=0.5,
        ylabel={\# execs},
        ymode=log,
        title={RS232-T600}, 
        x tick label style={rotate=45,anchor=east},
        x label style={at={(axis description cs:0.5,-0.2)},anchor=north},
        width=\columnwidth,      
        height=1.65in,      
        symbolic x coords={DFA, CFA, DCFA}, 
        legend style={
            at={(0,0)},
            anchor=south west,
            nodes={
                scale=0.6,
                transform shape
            }
        },
        xtick=data,     
        enlargelimits=0.25,
        ]
    \addplot+[pattern=north east lines, pattern color=.] coordinates {(DFA,79) (CFA,75) (DCFA,78)};
    \addplot+[pattern=north west lines, pattern color=.] coordinates {(DFA,92) (CFA,53) (DCFA,92)};
    \addplot+[pattern=crosshatch, pattern color=.] coordinates {(DFA,78) (CFA,59) (DCFA,78)};
    \legend{$\tau=max/4$,$\tau=max/8$, $\tau=min$}
  \end{axis}
\end{tikzpicture}
\end{subfigure}%
\begin{subfigure}[t]{0.325\textwidth}
\begin{tikzpicture} 
  \begin{axis}[      
        ybar=0.5,
        ylabel={\# execs},
        ymode=log,
        title={AE18\_core},
        x tick label style={rotate=45,anchor=east},
        x label style={at={(axis description cs:0.5,-0.2)},anchor=north},
        width=\columnwidth,      
        height=1.65in,      
        symbolic x coords={DFA, CFA, DCFA}, 
        legend style={
            at={(1,1)},
            anchor=north east,
            nodes={
                scale=0.6,
                transform shape,
            }
        },
        xtick=data,      
        enlargelimits=0.25,
        ]
    \addplot+[pattern=north east lines, pattern color=.] coordinates {(DFA,163) (CFA,94) (DCFA,71)};
    \addplot+[pattern=north west lines, pattern color=.] coordinates {(DFA,68) (CFA,64) (DCFA,74)};
    \addplot+[pattern=crosshatch, pattern color=.] coordinates {(DFA,68) (CFA,92) (DCFA,74)};
    \legend{$\tau=max/4$,$\tau=max/8$, $\tau=min$}
  \end{axis}
\end{tikzpicture}
\end{subfigure}
\begin{subfigure}[t]{0.325\textwidth}
\begin{tikzpicture} 
  \begin{axis}[      
        ybar=0.5,
        ylabel={\# execs},
        ymode=log,
        title={asyn\_fifo},
        x tick label style={rotate=45,anchor=east},
        x label style={at={(axis description cs:0.5,-0.2)},anchor=north},
        width=\columnwidth,      
        height=1.65in,      
        symbolic x coords={DFA, CFA, DCFA}, 
        legend style={
            at={(1,1)},
            anchor=north east,
            nodes={
                scale=0.6,
                transform shape,
            }
        },
        xtick=data,      
        enlargelimits=0.25,
        ]
    \addplot+[pattern=north east lines, pattern color=.] coordinates {(DFA,4703) (CFA,0) (DCFA,3065)};
    \addplot+[pattern=north west lines, pattern color=.] coordinates {(DFA,4703) (CFA,0) (DCFA,3065)};
    \addplot+[pattern=crosshatch, pattern color=.] coordinates {(DFA,3065) (CFA,0) (DCFA,3065)};
    \legend{$\tau=max/4$,$\tau=max/8$, $\tau=min$}
  \end{axis}
\end{tikzpicture}
\end{subfigure}
\caption{Comparison of number of executions (\# execs, logarithmic scale) with respect to threshold values ($\tau$) for different benchmarks and different configurations ($DFA$: data-flow only, $CFA$: control-flow only, $DCFA$: data-flow and control-flow) for {\bf \vgfa}. The Unidec design was excluded from the analysis since all tested threshold values resulted in a 24-hour timeout.}
\label{fig:threshold}

\end{figure*}
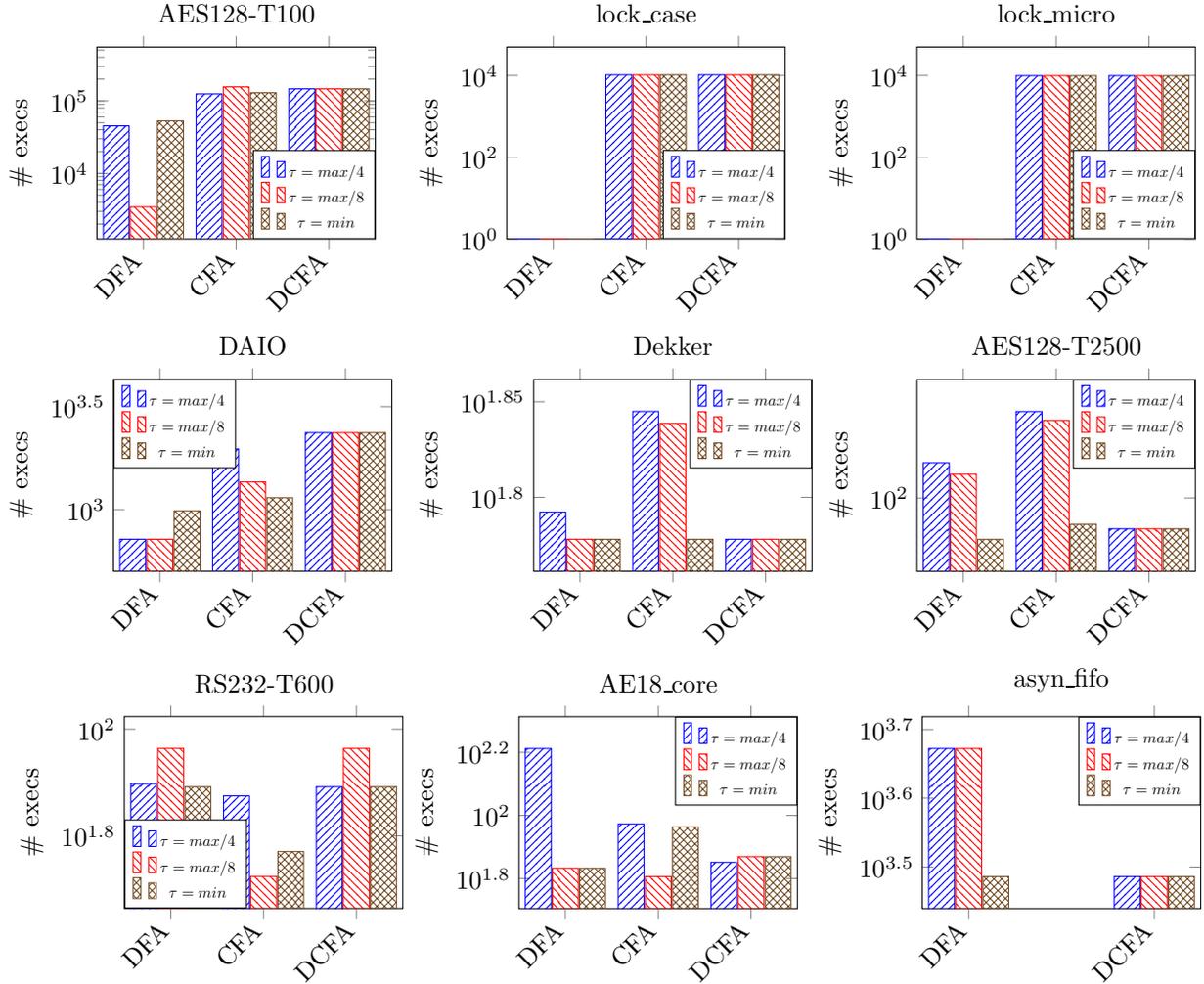

\subsubsection{Comparing Performances of \vgfa{}-static\_analysis to \vgfb{}-static\_analysis}
In this section, we use a threshold of $\tau=max/8$ and compare the effectivenss of {\vgfa} and {\vgfb} in terms of timing and total number of executions considering two open-source simulators, Icarus Verilog and Verilator.

\begin{figure*}[htb]
\begin{center}
\begin{tikzpicture} 
  \begin{axis}[      
        ybar=0.5,
        ymin=1, ymax=2000000,
        ylabel={\# execs},
        ymode=log,
        x tick label style={rotate=15,anchor=east},
        x label style={at={(axis description cs:0.5,-0.2)},anchor=north},
        width=\textwidth,      
        height=1.5in,      
        symbolic x coords={{AES128-T100, lock\_case, loc\_micro, DAIO, Dekker, Unidec, AES128-T2500, RS232-T600, AE18\_core, async\_fifo}}, 
        legend style={
            at={(0.5,1)},
            anchor=north east,
            legend columns=3,
            nodes={
                scale=0.8,
                transform shape
            }
        },
        xtick=data,
        enlarge y limits = 0,
        ]
        \addplot+[pattern=north east lines, pattern color=.] coordinates {(AES128-T100,3438) (lock\_case,3213) (loc\_micro,3167) (DAIO,718) (Dekker,56) (AES128-T2500,57) (RS232-T600,92) (AE18\_core,54) (async\_fifo, 3065)};
        \addplot+[pattern=north east lines, pattern color=.] coordinates {(AES128-T100,1) (lock\_case,9075) (loc\_micro,4799) (DAIO,828) (Dekker,105) (AES128-T2500,6858) (RS232-T600,133) (AE18\_core,217) (async\_fifo, 2000000)};
        \addplot+[pattern=north east lines, pattern color=.] coordinates {(AES128-T100,12158) (lock\_case,10137) (loc\_micro,10704) (DAIO,5380) (Dekker,23) (AES128-T2500,14702) (RS232-T600,305299) (AE18\_core,2063) (async\_fifo, 2000000)};
        \legend{\vgfa, \vgfb, HW-Fuzz}
  \end{axis}
\end{tikzpicture}
\caption{Comparison of total number of executions for \vgfa{}, \vgfb{}, and HW-Fuzz, Threshold $\tau=max/8$. Design {\bf Unidec} is not listed because it times out for all three techniques.}
\label{fig:execs_comp}
\end{center}
\end{figure*}
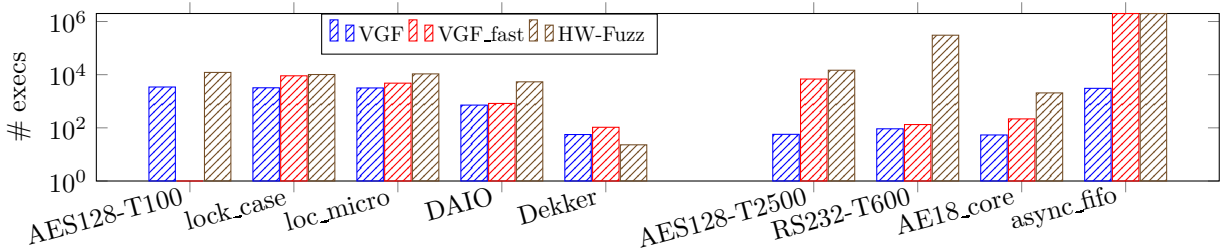

\paragraph{Number executions comparison}
We show the average number of executions required to trigger an assertion in Figure \ref{fig:execs_comp}.
Both \vgfa{} and \vgfb{} surpass HW-Fuzz across all benchmarks, with the exception of Dekker and, for \vgfb{}, AES128-T100. 
Moreover, \vgfa{} consistently outperforms \vgfb{}. This performance differential arises from Verilator's inherent optimization strategies when generating software models.

However, a constraint emerges from Verilator's inability to handle multi-clock domain vulnerabilities. As illustrated by the async\_fifo design, which incorporates both fast and slow clock domains, \vgfa{} can successfully identify the vulnerability. In contrast, both \vgfb{} and HW-Fuzz are unable to trigger this fault in a \SI{24}{\hour} time period. This is because Verilator is unable to properly handle proper timing on this type of design. It should be stressed that \emph{modern logic design makes use of multiple clock domains, which are used to drive multiple peripherals at the same time}. As such, we question the usability of Verilator for finding bugs in modern designs.

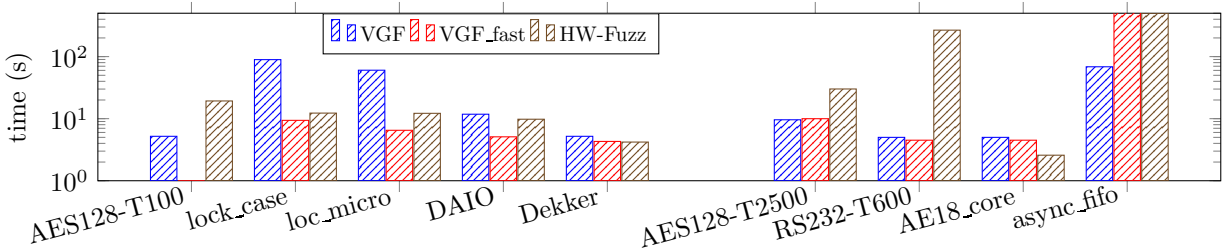
\begin{figure*}[htb]
\begin{center}
\begin{tikzpicture} 
  \begin{axis}[      
        ybar=0.5,
        ymin=1, ymax=500,
        ylabel={time (s)},
        ymode=log,
        x tick label style={rotate=15,anchor=east},
        x label style={at={(axis description cs:0.5,-0.2)},anchor=north},
        width=\textwidth,      
        height=1.5in,      
        symbolic x coords={{AES128-T100, lock\_case, loc\_micro, DAIO, Dekker, Unidec, AES128-T2500, RS232-T600, AE18\_core, async\_fifo}}, 
        legend style={
            at={(0.5,1)},
            anchor=north east,
            legend columns=3,
            nodes={
                scale=0.8,
                transform shape
            }
        },
        xtick=data,
        enlarge y limits = 0,
        ]
        \addplot+[pattern=north east lines, pattern color=.] coordinates {(AES128-T100,5.2) (lock\_case,89.4) (loc\_micro,60.4) (DAIO,11.8) (Dekker,5.2) (AES128-T2500,9.6) (RS232-T600,5.0) (AE18\_core,5.0) (async\_fifo, 68.6)};
        \addplot+[pattern=north east lines, pattern color=.] coordinates {(AES128-T100,1) (lock\_case,9.4) (loc\_micro,6.5) (DAIO,5.1) (Dekker,4.3) (AES128-T2500,9.99) (RS232-T600,4.5) (AE18\_core,4.5) (async\_fifo, 1000)};
        \addplot+[pattern=north east lines, pattern color=.] coordinates {(AES128-T100,19.3) (lock\_case,12.3) (loc\_micro,12.2) (DAIO,9.8) (Dekker,4.2) (AES128-T2500,30.1) (RS232-T600,267.04) (AE18\_core,2.58) (async\_fifo, 500)};
        \legend{\vgfa{}, \vgfb{}, HW-Fuzz}
  \end{axis}
\end{tikzpicture}
\caption{Comparison of average timing for \vgfa{}, \vgfb{} and HW-Fuzz. Threshold $\tau=max/8$. Design {\bf Unidec} is not listed because it times out for all three techniques.}
\label{fig:time_comp}
\end{center}
\end{figure*}

\paragraph{Timing comparison}
Figure \ref{fig:time_comp} presents the average time required for \vgfa{}, \vgfb{}, and HW-Fuzz to trigger the assertions. 
While \vgfa{} generally exhibits poorer time performance compared to \vgfb{} and HW-Fuzz, an exception is noted in the case of the async\_fifo design. This design presents a multi-clock domain vulnerability that Verilator is ill-equipped to handle. 
The performance discrepancy can be attributed to Verilator's compilation of HDL into an optimized and thread-partitioned model encapsulated in a C++/SystemC module. This results in a significant speed advantage over interpretive Verilog simulators such as Icarus Verilog.

Additionally, \vgfb{} demonstrates either similar (within one second) or superior performance over HW-Fuzz in six out of seven successfully evaluated designs. 
This performance edge may arise from \vgfb{}'s capability to capture more granular design change information by continually monitoring each signal's value alterations.

\subsection{Effect of Threshold Value}
\label{sec:threshold}

The number of execution cycles AFL needs to detect a fault in a design is subject to variations based on the selected signals. 
For instance, the outcomes of distinct signals from the identical {\tt AES128-T100} testbench design that are being tracked are demonstrated in Table \ref{tb:exp_results_signal}. 
The number of execution cycles that AFL requires is influenced by diverse factors, such as the number of signals being traced. 

For example, the results of static analysis show that the signals $aes\_128.a4.v0$ and $aes\_128.a1.v0$ are less relevant to the target signal. 
Tracking only the signal \lstinline[style=vcode]{aes_128.a4.v0} yields better performance than tracking both signals, with an observed number of executions ($\# execs$) of 47k compared to 97k when tracking both. 
Therefore, our findings suggest that adding irrelevant signals exacerbates the poor performance of VGF.
Furthermore, the specific role of the signal that is being tracked plays a crucial role. 
For instance, compared with signal \lstinline[style=vcode]{aes_128.a4.v0}, the target signal has stronger dependency with signal  \lstinline[style=vcode]{aes_128.k0}, thus, when tracing signal \lstinline[style=vcode]{aes_128.k0}, $\# execs$ is only about {20k}, while it increases to approximately {47k} when tracking signals \lstinline[style=vcode]{aes_128.a4.v0}.

Selecting the appropriate signals to be tracked during the execution of VGF is a crucial step, as it is necessary to choose a set of signals that is rich enough to capture changes in the targeted signal, while avoiding irrelevant ones. 
Therefore, a precise signal set should be designed with careful consideration.
Insight into the signals that need to be tracked during the execution of VGF can be found in the section on signal selection method. 
However, it is crucial to establish a threshold value $\tau$ for the static analysis to filter out signals. 
Only signals whose minimum distance to the property signals $ PD \leq \tau$ will be considered and fed into the VGF harness. 
This ensures that the fuzzing process is optimized and targeted towards relevant signals.
For the judicious selection of a meaningful threshold, $\tau$, we first determine the maximum and minimum $PDs$, denoted as {\tt max} and {\tt min}, across all signals. 
Based on these PD values, $\tau$ is categorized into five distinct levels: {\tt \{max, max/2, max/4, max/8, min\}}.

On one hand, as illustrated in Figure \ref{fig:numsignals}, the number of selected signals changes with different threshold values $\tau$ when only considering data-flow dependency (the control-flow dependency result is similar and thus not listed).
Due to the consideration of signals with less dependency on the property signal when using a large threshold value $\tau$, the number of selected signals generally increases as $\tau$ increase across most of the benchmarks.
However, the selection of the appropriate value for $\tau$, which determines the set of signals to be considered by the VGF harness, can be customized based on the characteristics of the hardware design being analyzed.
In the case of designs with a small number of internal signals, like {\tt DAIO}, {\tt Dekker}, and {\tt Unidec}, the number of selected signals does not vary significantly for values of $\tau=max/4$.
In contrast, hardware designs with a higher complexity level, involving a larger number of internal signals, such as {\tt AES128-T100/T2500} and {\tt AE18\_core}, may experience a significant rise in the number of selected signals, following an exponential pattern as $\tau$ decreases below $max/4$.
Thus, to avoid sacrificing the static analysis power, which may result from excessively small threshold values, we have set the maximum value of $\tau = max/4$.

On the other hand, in Figure \ref{fig:threshold}, the impact of different threshold values on the number of executions for various design implementations and different static analysis configurations is demonstrated.
It is evident that the effect of the threshold value on the number of executions differs among different design implementations and configurations.
For instance, in the case of the DFA-only configuration and $\tau = min$ scenario, the {\tt AES128-T2500} design requires fewer executions than {\tt AES128-T100}, which can be attributed to the presence of different Trojan types.

Therefore, we conclude that the threshold should be a range customized according to the specific implementations of the hardware design. 
Based on the discussions in this section, we limit the range of the $\tau$ to {\tt [min,max/4]} and choose the medium value $\tau=max/8$ in the following discussions. 



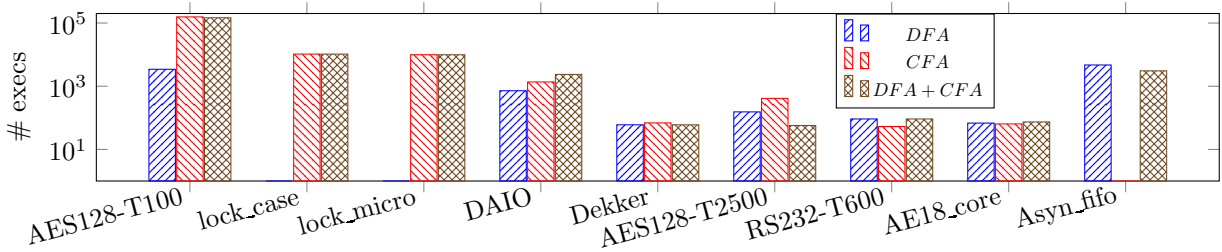
\begin{figure*}
\begin{center}
\begin{tikzpicture}
  \begin{axis}[      
        ybar=0.5,
        ymin=1, 
        ymax=200000,
        ylabel={\# execs},
        ymode=log,
        x tick label style={rotate=15,anchor=east},
        x label style={at={(axis description cs:0.5,-0.3)},anchor=north},
        width=\textwidth,
        height=1.5in,      
        symbolic x coords={AES128-T100, lock\_case, lock\_micro, DAIO, Dekker, AES128-T2500, RS232-T600, AE18\_core, Asyn\_fifo},
        legend style={
            at={(0.8,1)},
            anchor=north east,
            nodes={
                scale=0.7,
                transform shape
            }
        },
        xtick = {AES128-T100, lock\_case, lock\_micro, DAIO, Dekker, AES128-T2500, RS232-T600, AE18\_core, Asyn\_fifo},
        ]
    \addplot+[pattern=north east lines, pattern color=.] coordinates {(AES128-T100,3438) (lock\_case,1) (lock\_micro, 1) (DAIO,718) (Dekker,60) (AES128-T2500,154) (RS232-T600, 92) (AE18\_core, 68) (Asyn\_fifo, 4703)};
    \addplot+[pattern=north west lines, pattern color=.] coordinates {(AES128-T100,156284) (lock\_case,10408) (lock\_micro,9908) (DAIO,1362) (Dekker,69)  (AES128-T2500,412) (RS232-T600, 53) (AE18\_core, 64) (Asyn\_fifo, 1)};
    \addplot+[pattern=crosshatch, pattern color=.] coordinates {(AES128-T100,146846) (lock\_case,10408) (lock\_micro,9908) (DAIO,2362) (Dekker,60)  (AES128-T2500,57) (RS232-T600, 92) (AE18\_core, 74) (Asyn\_fifo, 3065)};
    \legend{$DFA$,$CFA$, $DFA+CFA$}
  \end{axis}
\end{tikzpicture}

\caption{Comparison of \# executions (\# execs) with respect to different configurations ($DFA$: data-flow only, $CFA$: control-flow only, $DFA+CFA$: data-flow and control-flow) for different benchmarks for {\bf \vgfa}, Threshold $\tau=max/8$. The \textbf{Unidec} design was excluded from the analysis since all configurations resulted in a 24-hour timeout.}
\label{fig:config}
\end{center}
\end{figure*}

\subsection{Effect of Configuration}
Based on the discussion in Section \ref{sec:threshold}, we use a threshold of $\tau=max/8$. 
Figure \ref{fig:config} shows that some designs, such as {\tt lock\_case} and {\tt lock\_micro}, have only control-flow dependencies, while some designs, such as {\tt AES128-T100} and {\tt DAIO}, have better results with only data-flow configuration. 
On the other hand, designs such as {\tt Dekker} and {\tt AES128-T2500} have better results with the combination of data-flow and control-flow configurations.

Different configurations can have a significant impact on the performance of VGF, as seen in the case of the {\tt AES128-T100} design, where the DFA-only configuration resulted in about $45\times$ and $42\times$ fewer $\# execs$ compared to CFA and CFA+DFA configurations, respectively. 
This is because designs that are dominated by data-flow paths perform better under the DFA configuration, whereas designs that are more control-flow dependent benefit from the CFA configuration. The DFA+CFA configuration performs best when both control-flow and data-flow are balanced in the design.

\subsection{Effect of Trimming}
One of the steps AFL performs is trimming. Trimming is a mutation step where AFL tries to shrink the size of the testcases it uses. The goal of is to reduce the complexity of the testcases which results in faster execution. According to \cite{fioraldi2023dissafl}, the trimming mutation step in AFL can cause a bottleneck as it wastes time trying to shorten inputs that do not necessarily need to be shortened. In this section we compare the results of running VGF with AFL trimming enabled and disabled in order to observe which setting should be chosen while fuzzing a hardware design. The results can be see in Table \ref{tb:trim_results}. Comparing the two results, removing the trimming step either improves or does not effect the results depending on the testbench. Therefore, we can conclude from these results that trimming is, in fact, detrimental to fuzzing these hardware testbenches and should be disabled when running VGF.

\begin{table}[htb]
 
    \begin{center}
        \begin{tabularx}{\columnwidth}{X*{4}{c}}
        \toprule
         \multirow{2}{*}{\textbf{Design}}
         & \multicolumn{2}{c}{\textbf{With Trim}} &
         \multicolumn{2}{c}{\textbf{No Trim}} \\
          &
         \textbf{DFA} & 
         \textbf{CFA} & \textbf{DFA} & \textbf{CFA} \\
         \midrule
         AES128-T100 & 3438 & 156284 & 2101 & 41643  \\
         lock\_case & N/A & 10408 & N/A & 3213 \\
         lock\_micro  & N/A & 9908 & N/A & 3167 \\
         DAIO & 718 & 1362 & 888 & 1493 \\
         Dekker & 60 & 69 & 56 & 73 \\
         Unidec & TO  & TO & TO & TO \\
         AES128-T2500 & 381 & 556 & 154 & 412 \\
         RS232-T600 & 92 & 53 & 78 & 53\\
         AE18\_core & 54 & 87 & 68 & 64 \\
         asyn\_fifo & 17067 & N/A & 4703 & N/A \\
         \bottomrule
        \end{tabularx}
    \caption{Average number of execution cycles to trigger fault from \vgfa{} with and without AFL input trimming. Average of 5 test runs per design and approach. Threshold $\tau=max/8$.}
    \label{tb:trim_results}
    \end{center}
    \vspace{-0.8cm}
\end{table}

\subsection{Effect of Deterministic Mutations}
AFL has the option to use what is known as deterministic mutations. These are predetermined mutations that AFL performs. According to \cite{AFL-github} these mutations result in more efficient testcases, however it also notes that due to the deterministic steps being more computationally intensive, it can significantly slow down the speed at which AFL is able to fuzz. In this section, we will compare VGF's performance with the deterministic steps enabled and disabled. The results can be seen in Table \ref{tb:determ_results}. Included in the table is the number of bits of input each design requires. From this result, we are able to see that the larger designs experience a severe performance loss with the deterministic steps enabled. The smaller designs either suffer a small performance penalty, or gain a small performance boost. From these results we see that as the required input size increases, deterministic steps get exponentially worse and only small gains are to be had with smaller sized inputs. It should also be noted that any performance gain observed from enabling deterministic steps is outweighed by the performance gain from disabling trimming.

\begin{table}[htb]
    \centering
        \begin{tabularx}{\columnwidth}{Xccccc}
        \toprule
         \multirow{2}{*}{\textbf{Design}} & \multicolumn{2}{c}{\textbf{Disabled}} &  
         \multicolumn{2}{c}{\textbf{Enabled}} & \multirow{2}{0.25in}{\centering\textbf{Input Size}} \\
          & \textbf{DFA} & 
         \textbf{CFA} & \textbf{DFA} & \textbf{CFA} &  \\
         \midrule
         AES128-T100 & 3438 & 156284 & 318879 & 540761 & 256  \\
         lock\_case & N/A & 10408 & N/A & 8711 & 8 \\
         lock\_micro  & N/A & 9908 & N/A & 5185 & 8  \\
         DAIO & 718 & 1362 & 2193 & 2131 & 14 \\
         Dekker & 60 & 69 & 157 & 182 & 2 \\
         Unidec & --  & -- & -- & -- & 5 \\
         AES128-T2500 & 381 & 556 & 92006 & 100911 & 256 \\
         RS232-T600 & 92 & 53 & 120 & 85 & 10 \\
         AE18\_core & 54 & 87 & 321 & 284 & 30\\
         Asyn\_fifo & 4703 & N/A & 12464 & N/A & 8\\
         \bottomrule
        \end{tabularx}
    \caption{Average number of execution cycles to trigger fault in design from VGF with and without AFL deterministic mutations. Average of 5 test runs per design and approach.}
    \label{tb:determ_results}
    \vspace{-0.7cm}
\end{table}

%% file: background/background.tex
Existing hardware fuzzing methodologies can be divided into 1) fuzzing directly on a hardware system and 2) fuzzing hardware as software.
The first category involves the application of software fuzzing algorithms to hardware programs to identify design flaws. RFuzz \cite{10.1145/3240765.3240842}, a pioneering method in this field, introduced fuzzing to hardware description language (HDL) codes for the first time \cite{fu2021fuzzing}. As an early hardware fuzzer, RFuzz employs mux-toggle coverage as feedback for fuzzing hardware. DirectFuzz \cite{canakci2021directfuzz} expands upon this concept by allocating more mutation energy to test cases that achieve coverage levels near a manually chosen model, consequently reaching faster coverage. However, due to instrumentation overhead, the mux-toggle metric requires better scalability for larger designs, such as BOOM and CVA6. To address this limitation, DIFUZZRTL \cite{hur2021difuzzrtl} introduces a new coverage metric, control-register coverage, though this metric still overlooks numerous security-critical vulnerabilities. TheHuzz \cite{kande2022thehuzz} utilizes code coverage metrics (branch and condition coverage) as feedback information. Nonetheless, TheHuzz needs help efficiently verifying the design under test due to difficulties covering hard-to-reach hardware regions. Furthermore, since TheHuzz relies on an ISA, it cannot handle hardware that does not execute instructions, such as peripheral IP cores.

The second category converts hardware into hardware emulation using EDA tools, subsequently fuzzing it as software.
A notable instance of this approach is HW-Fuzz \cite{trippel2022fuzzing}, which translates hardware design into a software model and employs coverage-guided software fuzzers for hardware fuzzing. However, the software model lacks support for certain hardware constructs, including latches and floating wires \cite{chen2023hypfuzz}. The equivalence between software and hardware conversions necessitates further verification \cite{sadeghi2021organizing}, and designing a comprehensive assertion system for the targeted hardware poses a significant challenge \cite{fu2021fuzzing} HyPFuzz \cite{chen2023hypfuzz}, a recent innovation, amalgamates fuzzing and formal verification techniques to validate large-scale processor designs. This method supports widely-used HDLs such as Verilog and SystemVerilog, addressing some of the limitations encountered by earlier hardware fuzzers.

Guided fuzzing in the software domain includes generation of a protocol state machine for fuzzing network servers \cite{Nat22}, 
tracking the values of state variables in protocol implementations \cite{BBM22},
using likely invariants to reward failure inducing inputs \cite{FDB21},  
and using static analysis to measure the distance to some target location of interest \cite{CXL18}.
These approaches are complementary to our property guided approach. It should further be noted that identifying signals of interest on software models of hardware presents challenges, as current software models perform aggresive optimizations on the design for speed purposes.

%% file: conclusions/conclusions.tex
This paper presented a state-based coverage metric for hardware fuzzing named VGF. 
We presented a sample implementation of VGF which is simulator and HDL agnostic. Our fuzzing strategy is capable of extracting state information from a design regardless of the microarchitectural choices made by designers. Additionally, we perform static analysis on designs to extract data-flow and control-flow information to decide which state information is most relevant to the fuzzer. Our strategy shows considerable improvement in fuzzing and coverage detection over previous hardware fuzzing approaches. We believe our coverage metric can be transparently applied to previous hardware fuzzing attempts, also improving their state coverage maps. We also present a faster solution for VGF that trades off fuzzing efficiency for raw speed. Because of limitations on the simulation interfaces we used in our implementation, we see some degradation in the overhead of the approach, and thus call upon improving these interfaces or providing dedicated interfaces as part of the core language standards to aid the collection of these metrics. Further research in this area will focus on providing these avenues.

%% file: discussion/discussion.tex
In this section we discuss the effects of the compression function used by the harness on the fuzzing process, as well as the effects of the selected signals on the design and the fitness function used by AFL.


\subsection{AFL Fitness Function}
AFL internally gives preferences to inputs based on the size of the input and the time it took the fuzzing target to process that input and terminate execution. That is, for an input bitstring $\vec{i}$, AFL computes the score $s = |\vec{i}| \times t_{\vec{i}}$, where $|\vec{i}|$ is the length size of the input and $t_{\vec{i}}$ is the time of execution. Then, for every bucket $b \in \mathbb{N}_n$ where $b$ has a non-zero value, AFL checks if $b$ has not been populated on a previous iteration. If the bucket has not been populated it is assigned a copy of the input $\vec{i}$ as well as the record of its score, otherwise AFL compares the recorded score with the newly computed one. If the newly computed score is \emph{smaller} than the previously recorded one, then the stored input is replaced with the new input vector for that bucket and the score is updated. This replacement is done regardless of the number of buckets, or control-flow paths, that the input bitstring populated.

This process implies that AFL attempts to minimize both execution runtime \emph{and} input size. This type of reward function has an unfortunate side effect on our fuzzing scheme. The more an input causes a signal to change, the longer the simulation time. This is because the simulator must perform two basic actions: it must propagate any side effects of the signal changing to the rest of the design, and if the signal is one of the ones we are tracking it must execute the callback function declared. The higher the number of simulation events, the longer the amount of time spent on the simulation. This is detrimental when attempting to discover the amount of state changes in the hardware as it is simulated given that the current fitness function will indirectly prioritize a small number of state changes.
As a result, we tested different functions in AFL that are agnostic to input size and execution time. The following functions were added to AFL

\subsubsection{Geometric Mean $\sqrt[n]{\prod a_i}$} The geometric mean of all populated bucket $n$ values $a_i$ is returned as the score. AFL keeps the highest score. This function rewards inputs that maximize the rate of state changes. Increasing the rate of state changes will increase the odds of a new control-flow path being discovered.

\subsubsection{Arithmetic Mean $\sum a_i /n$} The average of all populated bucket $n$ values $a_i$ is returned as the score. AFL keeps the highest score. This kind of fitness function emphasizes an equal distribution of state changes across all control-flow paths with the goal of ensuring all control-flow paths found are explored which should help maximize the odds of discovering a child control-flow path.

\subsubsection{Number of Buckets $n$} The total number of populated buckets $n$ is counted and returned as the score. AFL keeps the highest score. An input bitstring is rewarded for maximizing the number of buckets populated. The goal of rewarding the maximization of the number of populated buckets is to reward discovery of new buckets and, therefore, the discovery of new control-flow paths to promote increased coverage of the control-flow map.  

\subsubsection{Total Bucket Value $\sum n$} The total of all populated buckets $n$ is returned as the score. AFL keeps the highest score. This method rewards an input bitstring for maximizing the number of state changes that will speed up the rate at which a new control-flow path is found.


\subsubsection{Comparison of Fitness Functions}
Trials for the different compression functions were run and are recorded on Table \ref{tb:reward_functions}. As expected, default AFL fitness function yields worse performance, as it trends to prioritize small inputs which yield little in terms of signal changes.

\begin{table}[h]
    \begin{center}
        \begin{tabularx}{\columnwidth}{X *{5}{c}}
        \toprule
         \textbf{Design} & \textbf{AFL} & 
         $ \sqrt[n]{\prod a_i}$ & $\sum a_i/n $ & $n$ & $\sum a_i$ \\
         \midrule
         AES128-T100 & 1546 & 3938 & 2101 & 4227 & 5773  \\
         lock\_case & 13073 & 3918 & 13284 & 10705 & 3213 \\
         lock\_micro  & 21210 & 20679 & 3167 & 15134 & 13969  \\
         DAIO & 2905 & 1516 & 1418 & 1968 & 888 \\
         Dekker & 101 & 58 & 116 & 89 & 109 \\
         Unidec & -- & -- & -- & -- & -- \\
         AES128-T2500 & 787 & 309 & 291 & 156 & 259 \\
         RS232-T600 & 92 & 95 & 86 & 90 & 78 \\
         AE18\_core & 234 & 64 & 122 & 178 & 126 \\
         Asyn\_fifo & 3065 & 6973 & 13359 & 4762 & 10964 \\
         \bottomrule \\
        \end{tabularx} 
    \caption{Comparison of results of AFL's stock fitness function and modified versions. $n$ refers to the number of populated buckets and $a_i$ the non-zero bucket values. Fuzzing performed with the DFA configuration and trimming disabled. No data is given for when the fuzzer is unable to find faults within a 24 hour window.}
    \label{tb:reward_functions}
    \end{center}
\end{table}

Interestingly, two benchmarks go against this norm, with \texttt{AES128-T100} showing significantly better performance, and \texttt{RS232-T600} which shows nominally equal performance. However, the former design is very data driven. That is, being a cryptographic core, very small changes on the input produce rather drastic changes in the intermediates. As such, with the newly introduced functions, AFL is more prone to give priority to larger inputs which produce a wider array of changes in the internal state of the core. The latter core is relatively small and its activation condition requires a specific sequence to be added to the core. The complexity of the sequence is rather trivial for AFL to find with any reward function.

\subsection{Compression Function}
We also experimented with the compression function used by the VGF harness to populate the AFL control-flow hashmap. We stipulated that simulator accuracy and runtime would affect the performance of the fuzzer when choosing different options for our harness. We discuss these functions below.

\subsubsection{Compress Values}
A unique numeric signal identification code $c$ is given to every signal being tracked. Whenever a change in one of these signals occurs, the signal identification code $c$ is concatenated with the value the signal changed to $s_n$. Then, the operation $\{c, s_n\} \oplus \{c, s_{n-1}\}/2$ is performed, where $\{\circ, \circ\}$ is the bitwise concatenation operation, and $s_{n-1}$ is the previous value of the signal. We then compress the result using BLAKE2b function with a \SI{16}{\bit} digest. The use of BLAKE2b is arbitrary in nature, only being employed for collision resiliency. This operation is similar to the instrumentation employed by AFL.

\subsubsection{Vectorize Values}
Whenever a tracked signal changes, the values of all tracked signals are concatenated. The result is then hashed into a \SI{16}{\bit} digest and used to populate the AFL hashmap. Curiously, this approach has some interesting side effects in some designs. If multiple signals change at the same time, then the same position in the hashmap will be incremented multiple times. This is because the compression function is called multiple times per round of execution. However, if the signals change at different times due to delays in assignments, then the way the hashmap is populated is slightly differently. This condition can occur through a delayed assignment using the SystemVerilog or VHDL delay specifications. As an example, consider the code in Listing \ref{lst:compression_vector_case_b}.




\begin{lstlisting}[style=vcode,caption={Continuous assignments to tracked signals with assignment to one signal delayed},label={lst:compression_vector_case_b}]
    assign signal_a = some_logic;
#1  assign signal_b = some_other_logic;
\end{lstlisting}

Signal \lstinline[style=vcode]{signal_a} changes and the callback function executes. It uses the new updated value of this signal and the \emph{non-updated} value for \lstinline[style=vcode]{signal_b}. Then, after one simulation timestep, the callback for \lstinline[style=vcode]{signal_b} is updated resulting in the callback function once again being executed. However, the vector constructed during this callback is different to the previously constructed one. Lest a collision during compression occurs, two buckets will be populated in the control-flow hash map.


\subsubsection{Effects of Compression Function}
During our experimentation, there was no noticeable variation on the results obtained by using either compression function. As per our premise, feedback to the fuzzer uses changes in the tracked signals $s\in\mathbb{T}$ after these have been bucketed using a compression function. AFL's mutation engine uses these buckets as way of determining coverage. However, AFL makes \emph{no assumption} as to the position of the populated buckets in its control-flow hashmap while internally accounting for collisions. As such, results see little variation regardless of which compression function is used in the VGF harness.

\subsection{Improving Runtime Performance}
As a consequence of our fuzzing model, we stop the simulator every time a new input needs to be scheduled. Furthermore, the simulation is also halted while the callback function is executed. This is due to the way VPI/DPI/VHPI/FLI callbacks are mandated to be handled by language standards and as a consequence, the simulators. This slows down the fuzzing process as code harness code is executed sequentially outside the simulator. As a way to possibily increase fuzzing speed while maintaining simulation fidelity, we propose the creation of a new interface for simulators to directly communicate with external tools at a high speed. Data could be asynchronously exported by the simulator into a queue to be exposed to external tools. Similarly, the simulator could provide interfaces to asynchronously receive data from external tools as well. A test harness in the simulator could then be used to deterministically schedule the inputs provided by the external tool. The asynchronous nature of this methodology combined with the accuracy of emulation should yield quicker fuzzing times while still maintaining accurate results. Unfortunately, this would require rearchitecting major simulators while still providing all functionality. We believe such an interface should be included as part of HDL standard and urge a discussion on how to best develop it.